\documentclass[prd,aps,twocolumn,10pt,showpacs,preprintnumbers,amsmath,amssymb,floatfix]{revtex4-2}
\usepackage{graphicx}
\usepackage[breaklinks,colorlinks=true,linkcolor=blue,citecolor=blue,urlcolor=blue]{hyperref}
\usepackage{bm}
\usepackage{physics}
\usepackage{mathtools}
\usepackage{amsmath}

\newcommand{\Lag}{\mathcal{L}}
\newcommand{\ord}[1]{\mathcal{O}\!\left(#1\right)}
\newcommand{\Slash}[1]{\ooalign{\hfil/\hfil\crcr$#1$}}

\begin{document}

\title{Generalized parton distributions of the nucleon in the chiral quark soliton model II:  \\
Chiral anomaly mediation in the polarized GPDs}

\author{Josuke Minamiguchi}
\email{minamiguchi@nt.phys.s.u-tokyo.ac.jp}
\affiliation{Department of Physics, The University of Tokyo, 
  7-3-1 Hongo, Bunkyo-ku, Tokyo 113-0033, Japan}

\author{Tomoya Uji}
\email{uji@nt.phys.s.u-tokyo.ac.jp}
\affiliation{Department of Physics, The University of Tokyo, 
  7-3-1 Hongo, Bunkyo-ku, Tokyo 113-0033, Japan}

\begin{abstract}
We calculate the polarized flavor-singlet generalized parton distributions (GPDs) $\widetilde H^{u+d}$ and $\widetilde E^{u+d}$ in the chiral quark soliton model at nonzero skewness and momentum transfer.  Results are obtained with and without a singlet pseudoscalar meson $\eta_0$, whose mass is generated by the chiral anomaly.  Relative to the pion-only model, the singlet field introduces a sizable contribution to $\widetilde E^{u+d}$ in the ERBL region but only a minor correction to $\widetilde H^{u+d}$.  An explicit meson-exchange term qualitatively accounts for this contrast and displays the anticipated singlet pole.  When the anomaly-induced mass is set to zero, the pole becomes massless, and the exchange term dominates $\widetilde E^{u+d}$ at small momentum transfer.  These findings demonstrate how anomalous singlet-meson dynamics leaves distinct signatures in the polarized GPDs of the nucleon.
\end{abstract}

\maketitle

\section{Introduction}\label{sec:introduction}

The chiral anomaly provides a fundamental connection between quark chirality and gluon topology in quantum chromodynamics (QCD).  The flavor-singlet axial current, $J_5^\mu$, is not conserved due to the anomaly and satisfies the anomalous Ward identity~\cite{Adler:1969gk, Bell:1969ts}.  The forward nucleon matrix element of $J_5^\mu$ defines the flavor-singlet axial charge, $\Delta\Sigma$, which enters the first moment of the polarized structure function, $g_1$~\cite{Kodaira:1979pa, Jaffe:1989jz}.  The unexpectedly small $\Delta\Sigma$ inferred from polarized deep-inelastic scattering gave rise to the proton spin crisis~\cite{EuropeanMuon:1987isl}.  This observation stimulated renewed interest in the role of the anomaly in the interpretation of quark and gluon helicity contributions~\cite{Altarelli:1988nr, Carlitz:1988ab, Jaffe:1989jz}, as well as in the connection between nucleon spin and gluon
topology~\cite{Veneziano:1989ei, Narison:1994hv, Hatta:2020ltd, Tarasov:2020cwl, Tarasov:2021yll, Bhattacharya:2022xxw, Bhattacharya:2023wvy, Bhattacharya:2024geo, Tarasov:2025mvn, Fukushima:2026sot, Minamiguchi:2026uaq}.  The anomalous Ward identity thus links the spin structure of the nucleon to the dynamics responsible for anomalous $U(1)_A$ breaking.

Off-forward matrix elements of $J_5^\mu$ reveal an anomaly manifestation in the flavor-singlet induced pseudoscalar form factor, $G_P^{(0)}(t)$~\cite{Shore:1991dv}.  Together with the singlet axial form factor, $G_A^{(0)}(t)$, this form factor is obtained from the first moments of the polarized generalized parton distributions (GPDs)~\cite{Ji:1996ek}, i.e., $G_A^{(0)}(t) = \int_{-1}^{1}dx\,\widetilde{H}^{u+d}(x, \xi, t)$ and $G_P^{(0)}(t) = \int_{-1}^{1}dx\,\widetilde{E}^{u+d}(x, \xi, t)$.  In the chiral limit, the absence of the chiral anomaly would imply $G_P^{(0)}(t) = - 4M_N^2G_A^{(0)}(t) / t$, a relation that reflects the exchange of a massless flavor-singlet Nambu--Goldstone boson~\cite{Bhattacharya:2022xxw}.  At the perturbative level, the chiral anomaly manifests itself as an infrared $1/t$ singularity, known as the anomaly pole, in calculations of the polarized Compton amplitude~\cite{Tarasov:2020cwl, Bhattacharya:2022xxw}.  This infrared contribution can be absorbed into the quark GPD, $\widetilde{E}^{u+d}$, consistently with QCD factorization at one loop~\cite{Bhattacharya:2023wvy}.  This absorption alone, however, does not explain the absence of a massless pole in the full flavor-singlet GPD.  A nonperturbative cancellation between the anomaly pole and the would-be flavor-singlet Nambu--Goldstone pole has been proposed, leaving an $\eta'$-exchange contribution proportional to $1/(t - m_{\eta'}^2)$~\cite{Tarasov:2021yll, Bhattacharya:2022xxw}.  This mechanism connects the pole structure of $\widetilde{E}^{u+d}$ to the anomalous generation of the $\eta'$ mass through nonperturbative gluon topology~\cite{Witten:1979vv, Veneziano:1979ec}.  A nonlocal generalization of the anomalous Ward identity has subsequently provided an operator-level formulation of this cancellation through relations among twist-two, twist-three, and twist-four GPDs~\cite{Bhattacharya:2024geo}.

These relations constrain the pole structure of $\widetilde{E}^{u+d}$, but do not by themselves determine the magnitude and full momentum-fraction dependence of the meson-exchange contribution in the nucleon.  A quantitative investigation therefore requires a nonperturbative description that connects meson dynamics to quark light-front correlations.  Low-energy chiral models have also been used to investigate the effects of QCD anomalies on gravitational form factors~\cite{Fujii:2025aip, Tanaka:2025pny, Kawaguchi:2026yae}.  For the polarized GPDs considered here, the chiral quark soliton model (CQSM)~\cite{Diakonov:1987ty, Reinhardt:1988fz, Alkofer:1994ph, Christov:1995vm, Diakonov:1997sj} provides a suitable framework.  The CQSM combines constituent quarks and meson fields in a large-$N_c$ framework that incorporates the chiral symmetry of QCD and its spontaneous breaking.  This field-theoretical description has been applied to parton distribution functions (PDFs)~\cite{Weigel:1996kw, Diakonov:1996sr, Weigel:1996jh, Diakonov:1997vc, Weiss:1997rt, Wakamatsu:1997en, Pobylitsa:1998tk, Wakamatsu:1998rx, Weigel:1999pc, Wakamatsu:1999nj, Takyi:2019ahv}, GPDs~\cite{Petrov:1998kf, Penttinen:1999th, Ossmann:2004bp, Wakamatsu:2005vk, Kim:2023pll, Kim:2024ibz, Kim:2025mol, Minamiguchi:2026qcb}, transverse-momentum-dependent distributions~\cite{Pobylitsa:2003ty, Wakamatsu:2009fn, Schweitzer:2012hh}, and generalized transverse-momentum-dependent distributions (GTMDs)~\cite{Minamiguchi:2026usn}.  In particular, the pion-exchange contribution, $1/(t - m_\pi^2)$, to $\widetilde{E}^{u-d}$ was shown to emerge from the long-distance pion field~\cite{Penttinen:1999th}.  This result motivates an analogous investigation of the flavor-singlet channel.  An extension of the CQSM that retains a flavor-singlet pseudoscalar field $\eta_0$ with an anomaly-generated mass has recently been developed~\cite{Minamiguchi:2026uaq}, providing the necessary framework for such a calculation.

In the CQSM, the numerical evaluation of the polarized flavor-singlet GPDs nevertheless presents a substantial challenge.  Unlike their isotriplet counterparts, $\widetilde{H}^{u+d}$ and $\widetilde{E}^{u+d}$ require corrections beyond the leading large-$N_c$ approximation~\cite{Penttinen:1999th}.  The computational cost is particularly high when sea-quark contributions are included at nonzero skewness, $\xi$, and momentum transfer, $t$.  Earlier CQSM studies investigated the flavor-singlet axial charge~\cite{Suh:2022atr} and axial form factor~\cite{Silva:2005fa} of the nucleon, as well as the forward singlet helicity distribution~\cite{Wakamatsu:1998rx}.  Polarized GPDs at nonzero skewness were also studied in a light-front formulation, but the calculations were restricted to the three-quark Fock sector and the DGLAP region~\cite{Lorce:2011dv}.  These studies therefore left the ERBL region unexplored in the polarized flavor-singlet sector.  Access to this region is essential for understanding the singlet meson-exchange contribution.  Recently, one of the authors developed a method that substantially reduces the numerical cost of the spectral sums at nonzero $\xi$ and $t$~\cite{Minamiguchi:2026qcb}.  This development has enabled calculations of GPDs~\cite{Minamiguchi:2026qcb} and GTMDs~\cite{Minamiguchi:2026usn} with Dirac-sea contributions, opening the way to the polarized GPDs considered here.

In this work, we first calculate $\widetilde{H}^{u+d}$ and $\widetilde{E}^{u+d}$ at nonzero $\xi$ and $t$ in the conventional pion-only CQSM.  We then include the flavor-singlet pseudoscalar field and isolate the $\eta_0$-exchange contribution to $\widetilde{E}^{u+d}$.  The exchange term exhibits the expected singlet pole structure.  At the kinematics studied, the singlet field produces a substantial enhancement of $\widetilde{E}^{u+d}$, while the modification of $\widetilde{H}^{u+d}$ remains comparatively small.  This difference reflects the presence of the direct pseudoscalar exchange in $\widetilde{E}^{u+d}$, but not in $\widetilde{H}^{u+d}$.  The results provide a nonperturbative realization of the anomaly-related singlet pole within a nucleon model and quantify its impact on polarized GPDs beyond the forward axial charge.

\section{Theoretical Framework}\label{sec:theoretical}

In this section, we describe the CQSM with the flavor-singlet pseudoscalar field and present the model expressions for the polarized flavor-singlet GPDs.

\subsection{Chiral quark soliton model with \texorpdfstring{$\eta_0$}{eta0} meson}
\label{sec:model}

We start from the effective Lagrangian of the two-flavor CQSM extended to include the flavor-singlet pseudoscalar field, $\eta_0$~\cite{Minamiguchi:2026uaq},
\begin{equation}
    \Lag = \bar\psi\left(i\Slash{\partial} - M\mathcal{U}^{\gamma_5}\right)\psi - \frac{1}{2}m_0^2\eta_0^2\,,
    \label{eq:cqsm-lagrangian}
\end{equation}
where $\psi$ is the two-flavor constituent quark field and $M$ is the constituent quark mass generated by spontaneous chiral symmetry breaking.  The quark--meson interaction follows from the bosonized instanton-vacuum effective theory~\cite{Alkofer:1994ph, Diakonov:1997sj}.  The $U(2)$ chiral field combines the pion field, $\pi^a$, and $\eta_0$ as
\begin{equation}
    \mathcal{U}^{\gamma_5} = e^{i\gamma_5\eta_0/f_0}e^{i\gamma_5\tau^a\pi^a/f_\pi}\,,
\end{equation}
where $\tau^a$ are the Pauli matrices in flavor space and $f_\pi$ and $f_0$ are the decay constants of $\pi^a$ and $\eta_0$, respectively.  We restrict the present analysis to the two-flavor theory.  Connecting $\eta_0$ to the physical $\eta$ and $\eta'$ mesons would require a three-flavor extension that accounts for their mixing.

The last term in Eq.~\eqref{eq:cqsm-lagrangian} incorporates $U(1)_A$ breaking~\cite{DiVecchia:1980yfw, Witten:1980sp}, with
\begin{equation}
    m_0^2 = \frac{4\chi_{\mathrm{YM}}}{f_0^2}\,,
    \label{eq:m0-chi}
\end{equation}
where $\chi_{\mathrm{YM}}$ is the topological susceptibility of pure Yang--Mills theory.  We fix $m_0$ by the Witten--Veneziano relation~\cite{Witten:1979vv, Veneziano:1979ec} as described in Appendix~\ref{app:parameters}.

The quark functional integral yields
\begin{equation}
    S_{\mathrm{eff}}[U, \eta_0] = - iN_c\Tr\ln\!\left(i\Slash{\partial} - M\mathcal{U}^{\gamma_5}\right) - \frac{m_0^2}{2}\int\!d^4\!x\,\eta_0^2\,,
    \label{eq:effective-action}
\end{equation}
where $U = \exp(i\tau^a\pi^a/f_\pi)$ and $\Tr$ includes space-time, Dirac, and flavor indices; the color trace is displayed explicitly as $N_c$.

In the large-$N_c$ saddle-point approximation, the static baryon is described by a hedgehog background~\cite{Diakonov:1987ty},
\begin{equation}
    U_c^{\gamma_5}(\bm{r}) = \exp\left[i\gamma_5\bm\tau\cdot\bm{e}_r F(r)\right]\,,\qquad \eta_{0, c} = 0\,,
    \label{eq:hedgehog}
\end{equation}
with $r = |\bm{r}|$, $\bm{e}_r = \bm{r}/r$, and radial pion profile $F(r)$.  The subscript $c$ denotes the static configuration.  The quark eigenstates satisfy
\begin{equation}
    H_c\ket{n} = E_n\ket{n}\,,\qquad H_c = -i\gamma^0\gamma^k\partial_k + \gamma^0MU_c^{\gamma_5}\,,
    \label{eq:dirac-spectrum}
\end{equation}
with wave functions, $\Phi_n(\bm{r}) = \braket{\bm r}{n}$.  They have definite parity and grand spin, i.e., the sum of quark orbital angular momentum, spin, and isospin.  The occupied states (``occ'') comprise the discrete valence state and the negative-energy continuum; the remaining states are denoted by ``non''.  Minimizing $N_c\sum_{n\in\mathrm{occ}}E_n$ fixes $F(r)$ and defines the classical soliton mass, $M_{\mathrm{cl}}$.

Nucleon spin and isospin are obtained by quantizing the collective orientation $A(t)\in SU(2)$~\cite{Diakonov:1987ty, Wakamatsu:1990ud},
\begin{equation}
    U^{\gamma_5}(\bm{r}, t) = A(t)U_c^{\gamma_5}(\bm{r})A^\dagger(t)\,,\qquad A^\dagger\dot A = \frac{i}{2}\Omega^a\tau^a\,,
    \label{eq:collective-rotation}
\end{equation}
where $\Omega^a$ are the angular-velocity components.  The leading singlet response allowed by parity and hedgehog symmetry is~\cite{Minamiguchi:2026uaq}
\begin{equation}
    \eta_0(\bm{r}, t) = \eta(r)\Omega^a(t)e_r^a+\ord{\Omega^2}\,,
    \label{eq:eta-profile}
\end{equation}
with radial response function $\eta(r)$.  The same angular structure appears in Skyrme models~\cite{Jain:1987sz, Meissner:1988iv, Schechter:1999hg, Fukushima:2026sot}.

We express the quark field as $\psi(\bm{r}, t) = A(t)\psi_{\mathrm{rot}}(\bm{r}, t)$, where $\psi_{\mathrm{rot}}$ is the field in the rotating frame~\cite{Wakamatsu:1990ud, Wakamatsu:1998rx}.  Substituting this expression and the rotating pion field of Eq.~\eqref{eq:collective-rotation} into Eq.~\eqref{eq:cqsm-lagrangian} and expanding to first order in $\Omega$, we obtain the quark Hamiltonian in this frame~\cite{Minamiguchi:2026uaq},
\begin{equation}
    H_{\mathrm{rot}} = H_c + V_\Omega + V_\eta + \ord{\Omega^2}\,,
    \label{eq:rotating-dirac}
\end{equation}
where $V_\Omega = - iA^\dagger\dot A = \Omega^a\tau^a/2$ arises from the time derivative acting on $A(t)$, while $V_\eta = \eta_0\Gamma_\eta$ comes from expanding the singlet factor $e^{i\gamma_5\eta_0/f_0}$.  Here, we introduce
\begin{equation}
    \Gamma_\eta(\bm{r}) = i\gamma^0\gamma_5\frac{M}{f_0}U_c^{\gamma_5}(\bm{r})\,.
    \label{eq:Gamma-eta}
\end{equation}

The profile, $\eta(r)$, is determined by requiring the effective action to be stationary under variations of $\eta(r)$ at fixed $F(r)$ and $\Omega^a$.  Expanding the action to quadratic order in $\Omega$ gives the linear response equation~\cite{Minamiguchi:2026uaq}
\begin{equation}
    \int_0^\infty dr'\,{r'}^2\mathcal{K}(r, r')\eta(r') = j_\eta(r)\,.
    \label{eq:eta-response-equation}
\end{equation}
Here, we use the kernel,
\begin{equation}
    \mathcal K(r, r') = \left[\frac{4\pi}{3}m_0^2 - 2q_1(r)\right]\frac{\delta(r - r')}{r^2} - 2q_0(r, r')\,,
    \label{eq:eta-kernel}
\end{equation}
with
\begin{align}
    &q_0(r, r') = \frac{N_c}{3}\sum_{nm}'\Re\frac{y_{nm}^a(r)y_{mn}^a(r')}{E_m - E_n}\,,
    \label{eq:q0-kernel} \\
    &y_{mn}^a(r) = \int d\Omega_{\bm r}\,\Phi_m^\dagger(\bm r)e_r^a\Gamma_\eta(\bm r)\Phi_n(\bm r)\,,
    \label{eq:eta-vertex} \\
    &q_1(r) = - \frac{N_c}{3}\sum_{n\in\mathrm{occ}}\int d\Omega_{\bm{r}}\,\Phi_n^\dagger(\bm{r})\Gamma_{\eta^2}(\bm{r})\Phi_n(\bm{r})\,,
    \label{eq:q1-kernel} \\
    &\Gamma_{\eta^2} = - \gamma^0\frac{M}{2f_0^2}U_c^{\gamma_5}\,,
\end{align}
where $\sum_{nm}'=\sum_{n\in\mathrm{occ},\,m\in\mathrm{non}}$ and $d\Omega_{\bm r}$ is the solid-angle measure.  We also use the source term,
\begin{equation}
    j_\eta(r) = \frac{N_c}{3}\sum_{nm}'\frac{\Re\left[\mel{n}{\tau^a}{m}y_{mn}^a(r)\right]}{E_m - E_n}\,.
    \label{eq:eta-source}
\end{equation}
The Dirac-sea contributions are regularized as described in Appendix~\ref{app:parameters}.

Substituting the rotating fields of Eqs.~\eqref{eq:collective-rotation} and \eqref{eq:eta-profile} into the effective action in Eq.~\eqref{eq:effective-action} and evaluating the spatial and internal traces yields an effective action for the collective coordinate $A(t)$.  In the adiabatic approximation, this action defines the collective Lagrangian through
\begin{equation}
    S_{\mathrm{eff}} = \int dt\,L_{\mathrm{coll}}(A, \dot A)\,.
    \label{eq:collective-action}
\end{equation}
Keeping terms through second order in $\Omega$ and using the equation of motion for $\eta(r)$, Eq.~\eqref{eq:eta-response-equation}, gives~\cite{Minamiguchi:2026uaq}
\begin{equation}
    L_{\mathrm{coll}} = -M_{\mathrm{cl}}+\frac{I}{2}\Omega^a\Omega^a\,,\qquad I=I_\Omega+I_\eta\,,
    \label{eq:total-inertia}
\end{equation}
where $I_\Omega$ is the pion-only moment of inertia and $I_\eta$ is the correction from the singlet response,
\begin{align}
    I_\Omega &= \frac{N_c}{6}\sum_{nm}'\frac{\mel{n}{\tau^a}{m}\mel{m}{\tau^a}{n}}{E_m-E_n}\,,\label{eq:ordinary-inertia}\\
    I_\eta &= \int_0^\infty dr\,r^2\eta(r)j_\eta(r)\,.
    \label{eq:I-eta}
\end{align}
Collective quantization replaces $\Omega^a$ by $J^a/I$, with $J^a$ the nucleon spin operator, and gives the nucleon mass, $M_N = M_{\mathrm{cl}} + 3/(8I)$.  The regularization and numerical parameters are specified in Appendix~\ref{app:parameters}, and the large-$r$ treatment of $\eta(r)$ in Appendix~\ref{sec:tail}.

\subsection{Polarized GPDs in the model}\label{sec:gpd}

The polarized quark GPDs of the nucleon, $\widetilde{H}^q$ and $\widetilde{E}^q$, are defined by~\cite{Ji:1996ek, Diehl:2003ny}
\begin{align}
    &\int\frac{dz^-}{2\pi}\,e^{ix\bar P^+z^-}\mel**{P', S_3'}{\bar{\psi}_q\left(-\frac{z}{2}\right)\gamma^+\gamma_5\psi_q\left(\frac{z}{2}\right)}{P, S_3} \nonumber \\
    &\qquad = \frac{1}{\bar{P}^+}\bar{u}(P', S_3')\Bigl[\gamma^+\gamma_5\widetilde{H}^q(x, \xi, t) \nonumber \\
    &\qquad\qquad\qquad\qquad + \frac{\Delta^+\gamma_5}{2M_N}\widetilde{E}^q(x, \xi, t)\Bigr]u(P, S_3)\,,
    \label{eq:defGPD}
\end{align}
where the matrix element is evaluated at $z^+ = 0,\, \bm{z}_\perp = \bm{0}$.  Here, $\psi_q$ is the quark field of flavor $q$, and $z$ is the separation between the two quark fields.  For any four-vector $v^\mu$, the light-cone components are $v^\pm = (v^0 \pm v^3) / \sqrt{2}$ and $\bm{v}_\perp = (v^1, v^2)$.  The initial and final nucleon states have momenta $P$ and $P'$ and spins $S_3$ and $S_3'$, respectively.  The average nucleon momentum is $\bar{P} = (P + P') / 2$, while the momentum transfer is $\Delta = P' - P$, with $t = \Delta^2$.  The longitudinal momentum fraction is $x$, and the skewness is $\xi = - \Delta^+ / (2\bar{P}^+)$.  The nucleon momenta satisfy $P^2 = P'^2 = M_N^2$.  The corresponding Dirac spinor is denoted by $u(P, S_3)$ and normalized as $\bar{u}(P, S_3)u(P, S_3) = 2M_N$.  The gauge link between the two quark fields is unity in the model and is omitted.

In our model, explained in Sec.~\ref{sec:model}, the polarized GPDs are decomposed as
\begin{equation}
    \widetilde{H}^{u + d} = \widetilde{H}_\Omega + \widetilde{H}_\eta\,,
    \qquad \widetilde{E}^{u + d} = \widetilde{E}_\Omega + \widetilde{E}_\eta\,,
    \label{eq:split}
\end{equation}
where the subscripts $\Omega$ and $\eta$ distinguish the conventional rotational contributions and those involving the induced singlet vertex, respectively.  Applying the formalism of Ref.~\cite{Minamiguchi:2026qcb} with the singlet coupling of Ref.~\cite{Minamiguchi:2026uaq}, we obtain
\begin{widetext}
\begin{align}
    \widetilde{H}_\Omega(x, \xi, t) &= \frac{M_{\mathrm{cl}}N_c}{2I}\int\frac{dz^0}{2\pi}\,e^{iz^0xM_{\mathrm{cl}}}\left[\sum_{nm}^{(n, \neq)}\frac{e^{-iz^0E_n}}{E_n - E_m} - \sum_{nm}^{(m, \neq)}\frac{e^{-iz^0E_m}}{E_n - E_m} + iz^0\sum_{nm}^{(n, \neq)}e^{-iz^0E_n}\right]\mel{n}{\tau^b/2}{m}\mel{m}{\mathcal O_H^b(z^0)}{n}\,,
    \label{eq:H-omega}
\end{align}
\begin{align}
    \widetilde{H}_\eta(x, \xi, t) &= \frac{M_{\mathrm{cl}}N_c}{2I}\int\frac{dz^0}{2\pi}\,e^{iz^0xM_{\mathrm{cl}}}\left[\sum_{nm}^{(n, \neq)}\frac{e^{-iz^0E_n}}{E_n - E_m} -\sum_{nm}^{(m, \neq)}\frac{e^{-iz^0E_m}}{E_n - E_m} - iz^0\sum_{nm}^{(n, =)}e^{-iz^0E_n}\right]\mel{n}{Y^b}{m}\mel{m}{\mathcal{O}_H^b(z^0)}{n}\,,
    \label{eq:H-eta}
\end{align}
\begin{align}
    \widetilde{E}_\Omega(x, \xi, t) &= \frac{M_{\mathrm{cl}}^2N_c}{I\xi}\int\frac{dz^0}{2\pi}\,e^{iz^0xM_{\mathrm{cl}}}\left[\sum_{nm}^{(n, \neq)}\frac{e^{-iz^0E_n}}{E_n - E_m} - \sum_{nm}^{(m, \neq)}\frac{e^{-iz^0E_m}}{E_n - E_m} + iz^0\sum_{nm}^{(n, \neq)}e^{-iz^0E_n}\right]\mel{n}{\tau^b/2}{m}\mel{m}{\mathcal{O}^b(z^0)}{n}\,,
    \label{eq:E-omega}
\end{align}
\begin{align}
    \widetilde{E}_\eta(x, \xi, t) &= \frac{M_{\mathrm{cl}}^2N_c}{I\xi}\int\frac{dz^0}{2\pi}\,e^{iz^0xM_{\mathrm{cl}}}\left[\sum_{nm}^{(n, \neq)}\frac{e^{-iz^0E_n}}{E_n - E_m} - \sum_{nm}^{(m, \neq)}\frac{e^{-iz^0E_m}}{E_n-E_m} - iz^0\sum_{nm}^{(n, =)}e^{-iz^0E_n}\right]\mel{n}{Y^b}{m}\mel{m}{\mathcal{O}^b(z^0)}{n}\,.
    \label{eq:E-eta}
\end{align}
\end{widetext}
The abbreviated sums are defined by
\begin{equation}
    \sum_{nm}^{(n)} = \sum_{n\in\mathrm{occ}}\sum_{m = \mathrm{all}}\,, \qquad \sum_{nm}^{(m)} = \sum_{n = \mathrm{all}}\sum_{m\in\mathrm{occ}}\,,
    \label{eq:spectral-sums}
\end{equation}
where ``occ'' includes the discrete valence level and the negative-energy continuum, and ``all'' denotes the complete single-quark spectrum.  The additional labels $\neq$ and $=$ restrict the sums to $E_m\neq E_n$ and $E_m=E_n$, respectively.

The operators entering these expressions are
\begin{align}
    \mathcal{O}_H^b(z^0) &= \left(\delta^{b3} + 2M_{\mathrm{cl}}\xi\frac{\Delta_\perp^b}{\bm\Delta_\perp^2}\right)(1 + \gamma^0\gamma^3)\gamma_5 \nonumber \\
    &\qquad\times e^{-iz^0\hat{p}^3/2}e^{i\bm\Delta\cdot\hat{\bm{X}}}e^{-iz^0\hat{p}^3/2}\,, \label{eq:Oop-H} \\
    \mathcal{O}^b(z^0) &= \frac{\Delta_\perp^b}{\bm\Delta_\perp^2}(1 + \gamma^0\gamma^3)\gamma_5 \nonumber \\
    &\qquad\times e^{-iz^0\hat{p}^3/2}e^{i\bm\Delta\cdot\hat{\bm{X}}}e^{-iz^0\hat{p}^3/2}\,.
    \label{eq:Oop}
\end{align}
Here, $\hat{\bm{X}}$ and $\hat p^3$ are the three-dimensional position and longitudinal momentum operators, respectively.  We work in the Breit frame, where $\Delta^0 = 0$, and adopt the large-$N_c$ kinematics of Refs.~\cite{Ossmann:2004bp, Minamiguchi:2026qcb}.  Choosing the first transverse axis along $\bm{\Delta}_\perp$, we write
\begin{align}
    \bm\Delta &= \left(\sqrt{- t - 4M_{\mathrm{cl}}^2\xi^2}, \,0, \, - 2M_{\mathrm{cl}}\xi\right)\,, \nonumber\\
    \bm{\Delta}_\perp &= \left(\sqrt{- t - 4M_{\mathrm{cl}}^2\xi^2}, \, 0 \right)\,.
    \label{eq:kinematics}
\end{align}
For the index $b = 1, 2, 3$, the transverse components appearing in Eqs.~\eqref{eq:Oop-H} and \eqref{eq:Oop} are therefore $\Delta_\perp^b=\delta^{b1}\sqrt{-t-4M_{\mathrm{cl}}^2\xi^2}$, with $\bm\Delta_\perp^2=-t-4M_{\mathrm{cl}}^2\xi^2$.  The singlet vertex is specified by
\begin{equation}
    \mel{m}{Y^b}{n} = \int_0^\infty dr\,r^2\eta(r)y_{mn}^b(r)\,,
    \label{eq:rot-vertex}
\end{equation}
with $y_{mn}^b(r)$ defined in Eq.~\eqref{eq:eta-vertex}.  The numerical evaluation and regularization of the spectral sums in Eqs.~\eqref{eq:H-omega}--\eqref{eq:E-eta} are summarized in Appendix~\ref{app:parameters}.


\section{Numerical results}\label{sec:results}

\begin{figure}
    \centering
    \includegraphics[width=\columnwidth]{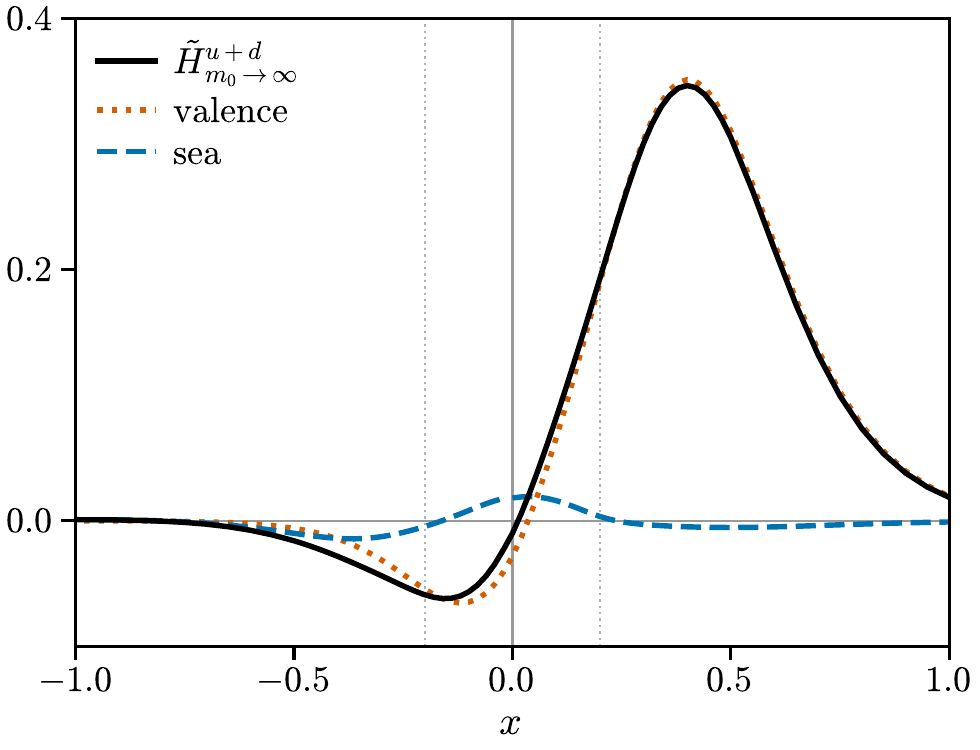}
    \caption{The polarized GPD $\widetilde H^{u+d}_{m_0\to\infty}$ (solid), with its valence (dotted) and Dirac-sea (dashed) contributions, at $\xi=0.20$ and $t=-0.30$~GeV$^2$.  The vertical dotted lines mark $x=\pm\xi$. The subscript $m_0\to\infty$ means $\widetilde{H}^{u + d}$ in the conventional pion-only CQSM.}
    \label{fig:pion-only-H}
\end{figure}

\begin{figure}
    \centering
    \includegraphics[width=\columnwidth]{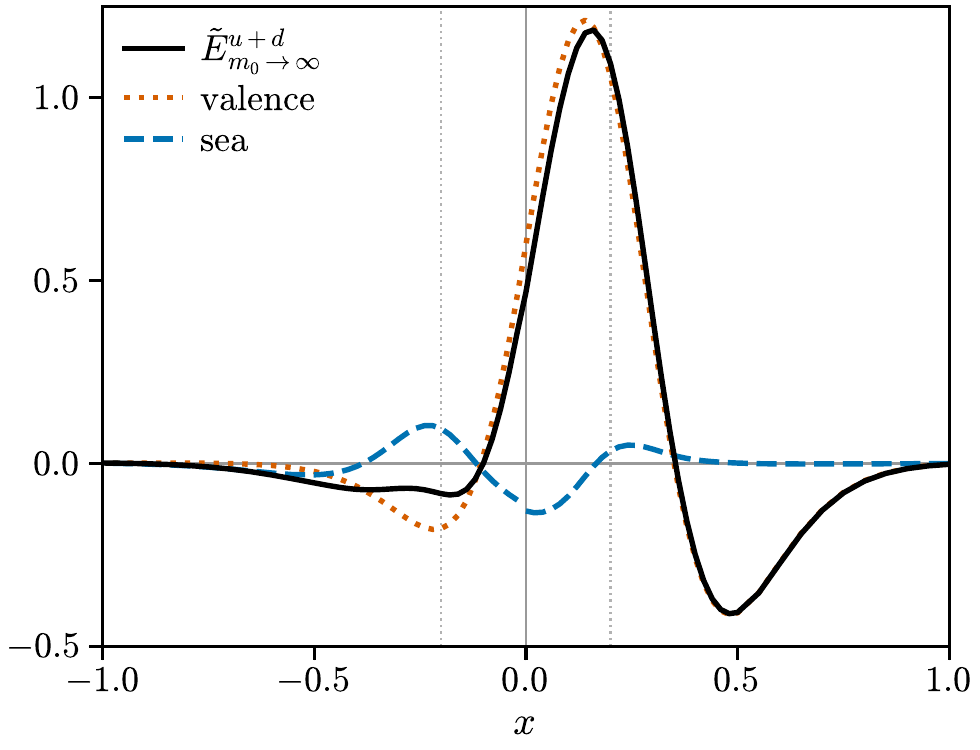}
    \caption{Same as Fig.~\ref{fig:pion-only-H}, but for $\widetilde E^{u+d}_{m_0\to\infty}$.}
    \label{fig:pion-only-E}
\end{figure}

We first present the results in the conventional pion-only CQSM, recovered in the limit $m_0\to\infty$, where the singlet field $\eta_0$ decouples.  Figures~\ref{fig:pion-only-H} and \ref{fig:pion-only-E} show $\widetilde H^{u+d}$ and $\widetilde E^{u+d}$, respectively, together with their valence and Dirac-sea contributions.  To our knowledge, these are the first numerical results for the polarized flavor-singlet GPDs in the CQSM at nonzero $\xi$ and $t$ with explicit summation over the regularized Dirac sea, covering both the DGLAP and ERBL regions.  They also provide the reference for identifying the effects of the singlet field.

We next include the singlet field with $m_0$ fixed by the Witten--Veneziano relation, i.e., $m_0 = 696$~MeV; see Appendix~\ref{app:parameters}.  Figures~\ref{fig:htilde} and \ref{fig:decomp} show the resulting GPDs at the same kinematics, decomposed into the $\Omega$ and $\eta$ contributions of Eq.~\eqref{eq:split}.  The $\Omega$ contributions remain close to the corresponding pion-only results, so the changes are mainly due to the explicit $\eta$ contributions.

\begin{figure}
    \centering
    \includegraphics[width=\columnwidth]{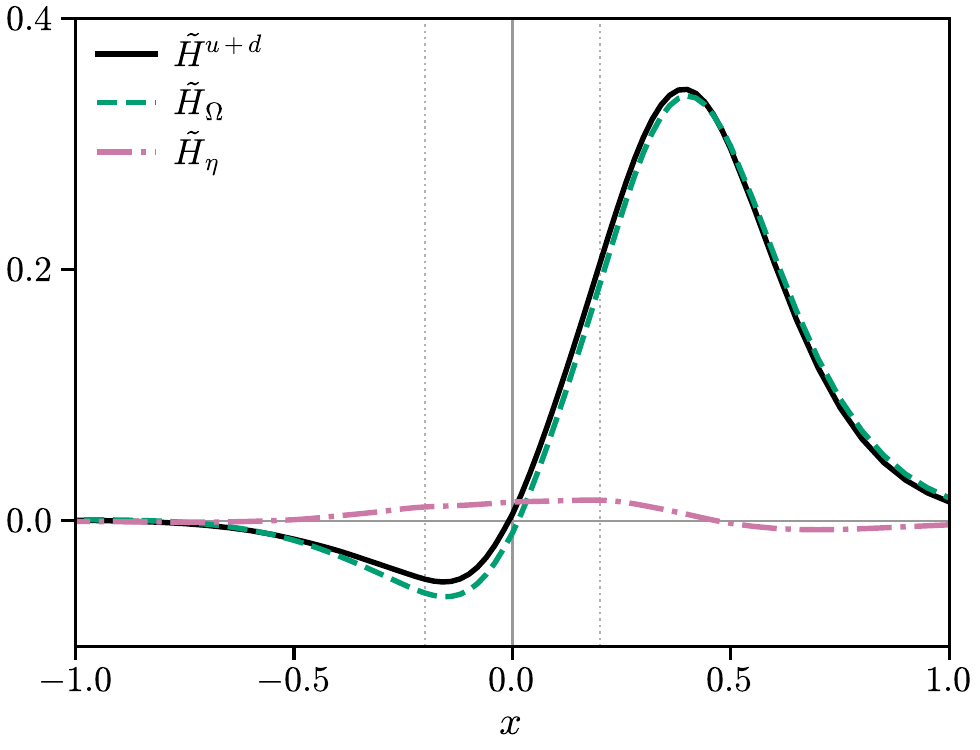}
    \caption{Decomposition of $\widetilde H^{u+d}$ (solid) into $\widetilde H_\Omega$ (dashed) and $\widetilde H_\eta$ (dash-dotted) at $\xi=0.20$ and $t=-0.30$~GeV$^2$ for $m_0=696$~MeV.  The vertical dotted lines mark $x=\pm\xi$.}
    \label{fig:htilde}
\end{figure}

\begin{figure}
    \centering
    \includegraphics[width=\columnwidth]{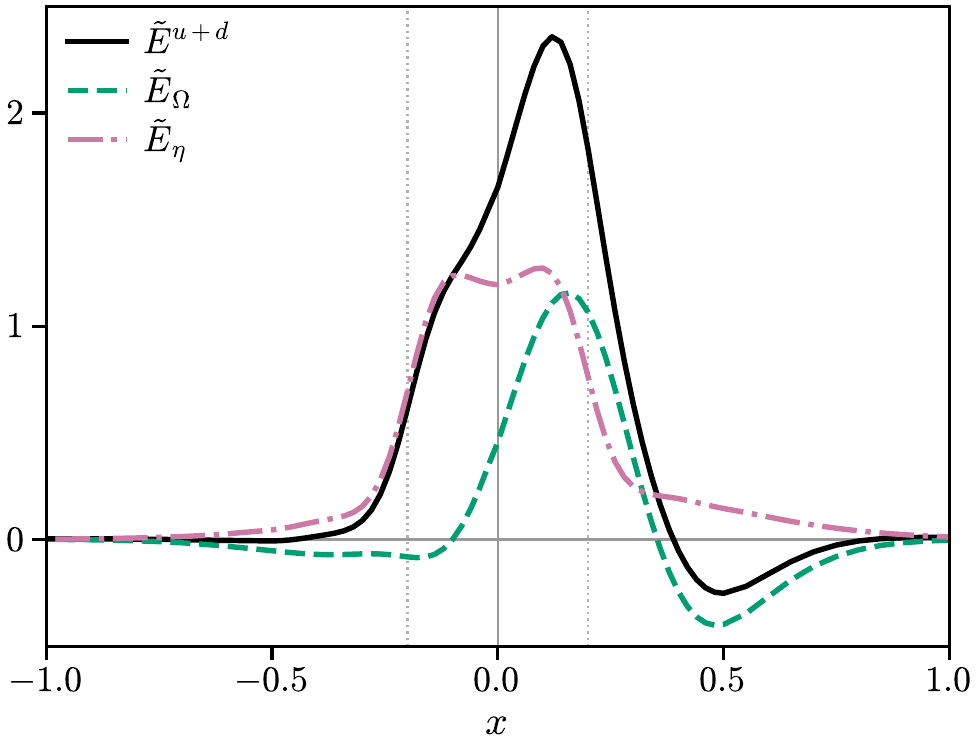}
    \caption{Same as Fig.~\ref{fig:htilde}, but for $\widetilde E^{u+d}$ (solid), $\widetilde E_\Omega$ (dashed), and $\widetilde E_\eta$ (dash-dotted).}
    \label{fig:decomp}
\end{figure}

For $\widetilde H^{u+d}$, the $\eta$ contribution is small, and the full distribution remains close to the pion-only result.  In contrast, $\widetilde E_\eta$ gives a large positive contribution in the ERBL region $|x|<\xi$, substantially enhancing $\widetilde E^{u+d}$ relative to the pion-only model.  The origin of this enhancement is examined in the next section by separating the $\eta_0$-exchange contribution from $\widetilde E_\eta$.

\section{Discussions}\label{sec:discussion}

To clarify the origin of the $\eta_0$-induced enhancement of $\widetilde E^{u+d}$, we isolate the contribution in which the soliton emits an $\eta_0$ meson that couples to the light-cone quark operator.  In this contribution, the singlet field is generated by the soliton, whereas its coupling to the operator is evaluated through a vacuum quark loop.  Accordingly, we replace the quark states in Eq.~\eqref{eq:E-eta} with free constituent-quark states and set $F = 0$ in $\Gamma_\eta$, while keeping the soliton-generated profile $\eta(r)$ unchanged.  By translational invariance of the vacuum, this response is the product of the Fourier component of $\eta(r)$ and the vacuum meson--operator vertex. Evaluating this vertex yields the meson distribution amplitude and gives the exchange contribution in the form derived in Appendix~\ref{app:exchange},
\begin{align}
    \widetilde{E}_\eta^{\mathrm{exch}}(x, \xi, t) &= G_P^{\mathrm{exch}}(t)\frac{\theta(\xi - |x|)}{2\xi}\phi_\eta\left(\frac{x + \xi}{2\xi}; t\right)\,.
    \label{eq:E-exch}
\end{align}
This expression has the same structure as the pion-exchange contribution to $\widetilde{E}^{u - d}$~\cite{Penttinen:1999th}.  Here, $G_P^{\mathrm{exch}}(t)$ is the induced pseudoscalar form factor specified below, $\theta$ is the step function, and $\phi_\eta(u; t)$ is the singlet-meson distribution amplitude at virtuality $t$, given in Eq.~\eqref{eq:phi-eta} and normalized as $\int_0^1du\,\phi_\eta(u;t)=1$.  The step function, $\theta(\xi - |x|)$, confines the exchange contribution to the ERBL region $|x|<\xi$.  We note that the anomaly-related pole terms discussed in Ref.~\cite{Bhattacharya:2023wvy} are not restricted to the ERBL region but also extend into the DGLAP region.  In contrast, the explicit $\eta_0$-exchange contribution isolated in our model has support only in $|x|<\xi$.  This constitutes a limitation of the present meson-exchange treatment, which does not address the more general anomaly-related pole structure in the DGLAP region.

The form factor, $G_P^{\mathrm{exch}}(t)$, is determined by
\begin{equation}
    G_P^{\mathrm{exch}}(t) = \frac{8\pi M_{\mathrm{cl}}^2}{I}f_A(t)\frac{\eta_1(\sqrt{-t})}{\sqrt{-t}}\,,
    \label{eq:GP-exch}
\end{equation}
with
\begin{equation}
    \eta_1(k) =\int_0^\infty dr\,r^2\mathsf{j}_1(kr)\eta(r)\,,
    \label{eq:eta1}
\end{equation}
where $\mathsf{j}_1$ is the spherical Bessel function of order one and $k=|\bm\Delta| = \sqrt{-t}$.  The axial coupling of $\eta_0$, defined by $J_5^\mu = f_A\partial^\mu\eta_0$, is $f_A(t) = 2f^2(t) / f_0$, with $f^2(t)$ of Eq.~\eqref{eq:Jsource}.  Both $f_A$ and $\phi_\eta$ are calculated from the regularized vacuum quark loop.  The exchange term is thus evaluated from the model without fitting its normalization or shape to $\widetilde E_\eta$.

The pole structure of $G_P^{\mathrm{exch}}(t)$ is encoded in the large-distance tail of the singlet profile.  As described in Appendix~\ref{sec:tail}, $\eta(r)$ approaches a form proportional to $(1+m_{\mathrm{pole}}r)e^{-m_{\mathrm{pole}}r}/r^2$, where $m_{\mathrm{pole}}$ is determined by Eq.~\eqref{eq:mpole}.  The Fourier transform of this tail generates a pole structure, i.e.,
\begin{equation}
    G_P^{\mathrm{exch}}(t) \sim \frac{1}{t - m_{\mathrm{pole}}^2}\,.
    \label{eq:pole-structure}
\end{equation}
The exponential decay rate therefore determines the pole position, while the asymptotic amplitude, together with the factors in Eq.~\eqref{eq:GP-exch}, determines its residue.  As $m_0\to0$, the tail becomes proportional to $1/r^2$, and the pole moves to $t=0$.

\begin{figure}
    \centering
    \includegraphics[width=\columnwidth]{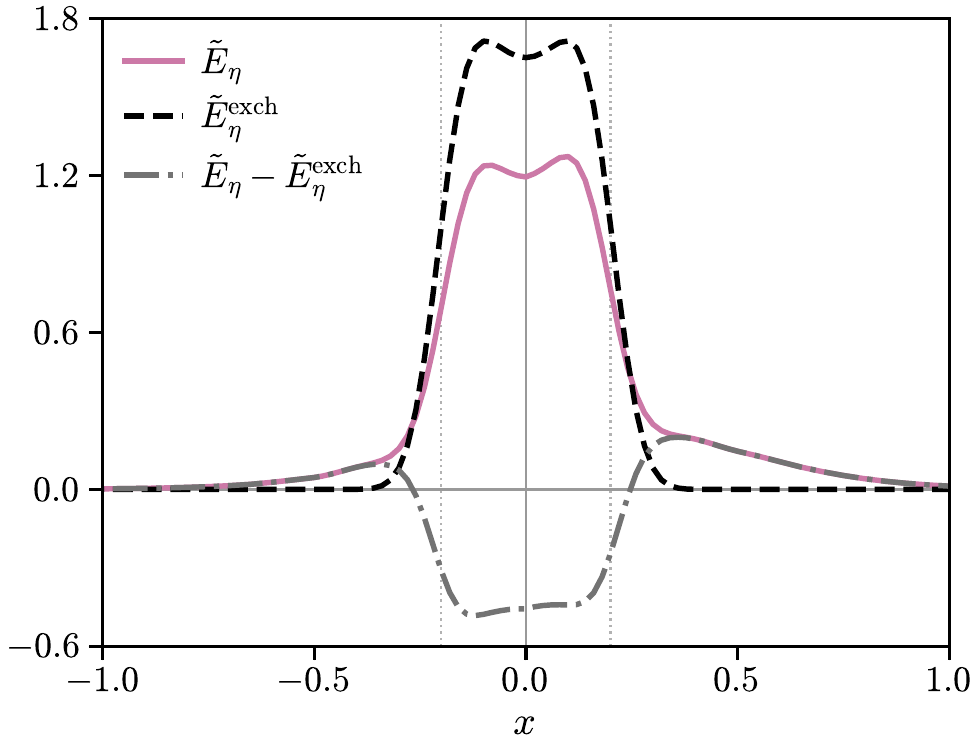}
    \caption{The singlet contribution $\widetilde E_\eta$ (solid), the $\eta_0$-exchange term $\widetilde E_\eta^{\mathrm{exch}}$ (dashed), and their difference (dash-dotted) at $\xi=0.20$ and $t=-0.30$~GeV$^2$ for $m_0=696$~MeV.  The exchange term is subjected to the same Gaussian smearing as the numerical GPD.  The vertical dotted lines mark $x=\pm\xi$.}
    \label{fig:exchange}
\end{figure}

Figure~\ref{fig:exchange} compares the singlet contribution $\widetilde{E}_\eta$ previously shown in Fig.~\ref{fig:decomp} with the derived exchange term $\widetilde E_\eta^{\mathrm{exch}}$, at the same parameters.  This comparison examines how much of the singlet-induced enhancement can be explained by $\eta_0$ exchange.  Although the exchange term in Eq.~\eqref{eq:E-exch} is confined to the ERBL region, the curve shown here is subjected to the same Gaussian smearing used in the numerical CQSM calculation.  This convolution smooths the behavior at $x = \pm\xi$ and spreads part of the ERBL-supported contribution into the DGLAP region.  The tails of the exchange curve outside $|x| < \xi$ therefore do not represent an intrinsic DGLAP contribution of $\eta_0$ exchange.

The exchange term produces a large positive contribution in the ERBL region, but overshoots $\widetilde E_\eta$ near the center, where the remainder $\widetilde E_\eta-\widetilde E_\eta^{\mathrm{exch}}$ is negative.  Thus, $\eta_0$ exchange provides a qualitative explanation for the enhancement of $\widetilde E^{u+d}$, although additional contributions are needed to reproduce the full momentum-fraction dependence at these kinematics.  A nonzero remainder also persists in the DGLAP region, showing that the distribution there cannot be explained solely by the smearing of the ERBL-supported exchange term.

To examine a regime in which the pole contribution is enhanced, we also consider $m_0=0$ at smaller $|t|$, reducing $\xi$ consistently with the kinematic restriction in Eq.~\eqref{eq:kinematics}.  In this limit, $m_{\mathrm{pole}}=0$ and the singlet profile falls as $1/r^2$ at large $r$.  Consequently, $\eta_1(k)\sim1/k$ for small $k$, and $G_P^{\mathrm{exch}}(t)$ grows as $1/(-t)$ toward $t=0$.  Moreover, the vacuum-loop distribution amplitude approaches $\phi_\eta(u;0)=1$, so the unsmeared exchange term becomes nearly constant across the ERBL region.

\begin{figure}
    \centering
    \includegraphics[width=\columnwidth]{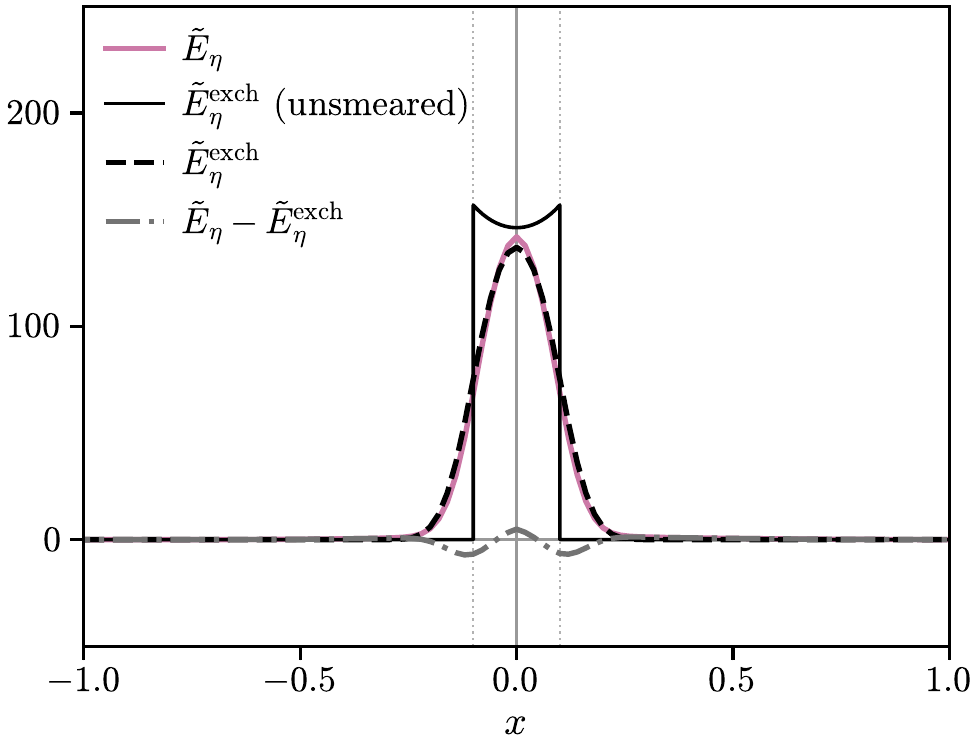}
    \caption{Same as Fig.~\ref{fig:exchange}, but for $m_0=0$ at $\xi=0.10$ and $t=-0.05$~GeV$^2$.  The thin solid line additionally shows the exchange term before Gaussian smearing.}
    \label{fig:exchange-m0}
\end{figure}

As shown in Fig.~\ref{fig:exchange-m0}, the smeared exchange term now closely reproduces $\widetilde E_\eta$.  The unsmeared contribution is confined to $|x|<\xi$ and has a nearly flat shape.  Smearing rounds this structure into a peak and generates tails beyond the ERBL boundaries.  Applying the same smearing to the derived exchange term reproduces both the peak and these tails in the numerical result.  The agreement therefore identifies an ERBL-supported meson-exchange contribution even though the plotted distribution extends into the DGLAP region.  Since the $\Omega$ contribution is also small at these kinematics, the exchange term accounts for almost all of $\widetilde E^{u+d}$.

The same vacuum-loop construction gives no exchange contribution to $\widetilde H^{u+d}$,
\begin{equation}
    \widetilde H_\eta^{\mathrm{exch}}=0\,,
    \label{eq:H-exch}
\end{equation}
as demonstrated in Appendix~\ref{app:exchange}.  The pseudoscalar-exchange structure contributes to $\widetilde{E}^{u+d}$ rather than $\widetilde{H}^{u+d}$.  Although the remaining singlet response of $\widetilde H^{u+d}$ need not vanish, it is numerically small, consistent with the weak modification observed in Sec.~\ref{sec:results}.

From the discussions above, our results show that the singlet field substantially enhances $\widetilde E^{u+d}$ while leaving $\widetilde H^{u+d}$ nearly unchanged.  This contrast is qualitatively explained by $\eta_0$ exchange, which contributes to $\widetilde{E}^{u+d}$ but not to $\widetilde{H}^{u+d}$.  Our calculation thus provides a nonperturbative realization, within the two-flavor CQSM, of the singlet meson pole in $\widetilde{E}^{u+d}$ anticipated from analyses of the chiral anomaly~\cite{Tarasov:2021yll, Bhattacharya:2022xxw, Bhattacharya:2024geo}.

\section{Conclusions}\label{sec:conclusion}
We have evaluated the polarized flavor-singlet GPDs $\widetilde H^{u+d}$ and $\widetilde E^{u+d}$ in the two-flavor CQSM.  The calculation covers both the DGLAP and ERBL regions at nonzero skewness and momentum transfer.  We obtained the distributions in the pion-only CQSM as a baseline and examined the changes produced by the flavor-singlet pseudoscalar field $\eta_0$, whose mass incorporates anomalous $U(1)_A$ breaking.

The numerical results reveal a pronounced sensitivity to the singlet field in $\widetilde E^{u+d}$, particularly in the ERBL region, in contrast to the weak response of $\widetilde H^{u+d}$.  Separating the $\eta_0$-exchange term identifies the meson pole responsible for this qualitative difference: the exchange contributes to $\widetilde E^{u+d}$, while its counterpart in $\widetilde H^{u+d}$ vanishes.  When the anomaly-induced mass is removed, the pole reaches $t=0$, and the exchange term closely reproduces the calculated $\widetilde E^{u+d}$ at small $|t|$.  These findings give concrete support, in a nonperturbative nucleon calculation, to the predicted connection between the chiral anomaly and singlet meson exchange in polarized GPDs~\cite{Tarasov:2021yll, Bhattacharya:2022xxw, Bhattacharya:2024geo}.

A natural next step is to investigate the analogous role of the trace anomaly in unpolarized GPDs.  Extending the framework to include scalar gluonic degrees of freedom could allow a direct examination of the proposed glueball-exchange contributions to unpolarized GPDs~\cite{Bhattacharya:2022xxw, Bhattacharya:2023wvy}.

\begin{acknowledgments}
The authors thank Kenji~Fukushima for useful discussions.  This work was carried out during the long-term workshop ``New Frontiers in QCD 2026'' at the Yukawa Institute for Theoretical Physics (YITP-T-26-01).  Discussions during the workshop were useful to complete this work.  The authors used Anthropic's Claude and OpenAI's ChatGPT to assist with the implementation and debugging of the numerical code and in the preparation of the manuscript.  This work was partially supported by IIW, WINGS Program, the University of Tokyo (J.M.) and by FoPM, WINGS Program, the University of Tokyo (T.U.).
\end{acknowledgments}

\appendix

\section{Numerical settings}\label{app:parameters}

We evaluate the double sums of Eqs.~\eqref{eq:H-omega}--\eqref{eq:E-eta} with the method of Ref.~\cite{Minamiguchi:2026qcb}.  With $\varpi = (x + \xi)M_{\mathrm{cl}} - E_n$, the integral over $z^0$ leaves the operators $e^{i\bm\Delta\cdot\hat{\bm{X}}}$ and $\delta(\varpi - \hat{p}^3)$, which are expanded as
\begin{align}
    &e^{i\bm\Delta\cdot\bm{r}} = \sum_{L = 0}^{L_{\max}}i^L(2L + 1)\mathsf{j}_L(|\bm\Delta|r)P_L(\bm{e}_{\bm\Delta}\cdot\bm{e}_r)\,,
    \label{eq:app-rayleigh} \\
    &\delta(\varpi - |\bm{p}|\bm{e}_p\cdot\bm{e}_3) = \sum_{\lambda = 0}^{\infty}\frac{2\lambda + 1}{2|\bm p|}P_\lambda\left(\frac{\varpi}{|\bm p|}\right) \nonumber \\
    &\qquad\times\theta(|\bm{p}| - |\varpi|)P_\lambda(\bm{e}_p\cdot\bm{e}_3)\,,
  \label{eq:app-lc}
\end{align}
where $\mathsf{j}_L$ are the spherical Bessel functions and $P_L$ are the Legendre polynomials.  In the Kahana--Ripka basis~\cite{Kahana:1984be}, these expansions reduce the sums over all quark states to radial integrals, without an interpolation formula and without a gradient expansion.  The vertex $Y^b$ is a Hermitian vector operator under the grand spin, like $\tau^b$, so this method applies unchanged.  We replace the delta function by a Gaussian~\cite{Diakonov:1997vc},
\begin{equation}
    \delta(\varpi)\to\frac{1}{\sqrt{\pi}\gamma M_{\mathrm{cl}}}e^{-\varpi^2 / (\gamma M_{\mathrm{cl}})^2}\,, \label{eq:app-smear}
\end{equation}
and show the distributions computed from the sum over the occupied states for $x\ge0$ and from the sum over the nonoccupied states for $x<0$~\cite{Minamiguchi:2026qcb}.

We regularize the quark loop of Eq.~\eqref{eq:effective-action} with the double Pauli--Villars subtraction~\cite{Kubota:1999hx}.  With $(\mu_0, \mu_1, \mu_2) = (M, \Lambda_1, \Lambda_2)$ and $(w_0, w_1, w_2) = (1, -c_1, -c_2)$, a quantity $\mathcal{Q}$ of the quark loop becomes
\begin{equation}
    \mathcal{Q}^{\mathrm{reg}} = \sum_{i = 0}^{2}w_i\left(\mathcal{Q}[\mu_i; F] - \mathcal{Q}[\mu_i; 0]\right)\,,
    \label{eq:Qreg}
\end{equation}
where $\mathcal{Q}[\mu; F]$ is computed with the quark mass $\mu$ in $H_c$ and in the $\eta$ vertex and with the profile $F$.  The conditions $\sum_{i = 1}^{2}c_i\Lambda_i^2 = M^2$ and $\sum_{i = 1}^{2}c_i\Lambda_i^4 = M^4$ fix $c_1$ and $c_2$.  The vacuum term has $F\equiv0$ and $\eta\equiv0$.  The same $F(r)$ and $\eta(r)$ are used at all three masses.  The mass term of $\eta_0$ is not subtracted, and the contribution of the valence state is not regularized~\cite{Ossmann:2004bp}.  Alternative subtraction schemes are discussed in Refs.~\cite{Weigel:1999pc, Takyi:2019ahv, Weigel:2021pbr}.

We work in the chiral limit with $M = 350$~MeV, $f_\pi = 93$~MeV, and $\langle\bar{q}q\rangle = - (286.5~\mathrm{MeV})^3$, which give $\Lambda_1 = 634.05$~MeV and $\Lambda_2 = 1505.53$~MeV.  We set $f_0 = f_\pi$ for simplicity.  For $m_0$ we take the Witten--Veneziano value at $N_f = 2$, $m_0^2 = \tfrac{2}{3}\left(m_\eta^2 + m_{\eta'}^2 - 2m_K^2\right) = (696~\mathrm{MeV})^2$~\cite{Witten:1979vv, Veneziano:1979ec}.

We use the box, the basis, and the self-consistent profile of Ref.~\cite{Minamiguchi:2026qcb}, with $\gamma = 0.08$ and $L_{\max} = 16$.  The soliton has $M_{\mathrm{cl}} = 1062$~MeV and $I_\Omega = 7.69\times10^{-3}$~MeV$^{-1}$.

\section{Tail of the singlet profile}\label{sec:tail}

The tail of $\eta(r)$ at large $r$ fixes the pole and the residue of the $\eta_0$ exchange term of Sec.~\ref{sec:discussion}.  We solve Eq.~\eqref{eq:eta-response-equation} in the spherical box of the quark basis.  Besides the contribution of the soliton, the computed source $j_\eta$ contains a structure near the boundary of the box.  For boxes of different sizes, this structure has the same shape as a function of the distance from the boundary.  We therefore regard this structure as an artifact of the box.  Reference~\cite{Minamiguchi:2026uaq} used the computed source as it was and replaced $\eta(r)$ at $r\ge4$~fm by a Yukawa form.  Here we remove the artifact from the source and impose the asymptotic form of $\eta(r)$ as follows.

We define the source-free interval $[r_A, r_c]$ as the interval outside the peak of $r^2|j_\eta(r)|$ in which
\begin{equation}
    r^2|j_\eta(r)| < \epsilon\,\max_{r'}r'^2|j_\eta(r')|
    \label{eq:source-free}
\end{equation}
holds without interruption.  The maximum is taken over the inner half of the box, and we use $\epsilon = 2\times10^{-4}$.  The radius $r_A$ is the point where the source of the soliton falls below the threshold, and $r_c$ is the point where the boundary structure exceeds the threshold.  The threshold is about ten times larger than the numerical noise of $r^2|j_\eta|$ in the interval.  We set $j_\eta = 0$ for $r > r_c$ and solve Eq.~\eqref{eq:eta-response-equation}.  We then replace the solution for $r>r_A$ by the free solution of $\eta_0$ in the vacuum, matched in value at $r_A$:
\begin{align}
    \eta(r) = \eta(r_A)\left(\frac{r_A}{r}\right)^2e^{-m_{\mathrm{pole}}(r-r_A)}\frac{1+m_{\mathrm{pole}}r}{1+m_{\mathrm{pole}}r_A}\,.
    \label{eq:eta-tail}
\end{align}
The mass $m_{\mathrm{pole}}$ is the pole of the $\eta_0$ propagator in the vacuum,
\begin{align}
    &f^2(m_{\mathrm{pole}}^2)\,m_{\mathrm{pole}}^2 = f_0^2\,m_0^2\,,
    \label{eq:mpole} \\
    &f^2(q^2) = \frac{N_c}{4\pi^2}\,J(q^2)\,,
    \label{eq:Jsource} \\
    &J(q^2) = -\sum_{i=0}^{2}w_i\,\mu_i^2\int_0^1du\,\ln\!\big[\mu_i^2-u(1-u)q^2\big]\,,
  \label{eq:J}
\end{align}
with $\mu_i$ and $w_i$ of Eq.~\eqref{eq:Qreg}.  The function $f^2(q^2)$ is the square of the momentum-dependent decay constant given by the vacuum quark loop, with $f^2(0)=f_\pi^2$.  The kinetic term of $\eta_0$ generated by the same loop has the coefficient $f^2(q^2)/f_0^2$, which equals unity at $q^2=0$ for $f_0=f_\pi$.  Because $f^2(q^2)>f_0^2$ for $0<q^2<4M^2$, the pole mass $m_{\mathrm{pole}}$ is smaller than $m_0$.  Equation~\eqref{eq:eta-tail} keeps only the pole of the propagator.  The contribution of the quark-antiquark continuum above $q^2=4M^2$ decays approximately as $e^{-2Mr}$ and is neglected.  The prescription does not depend on $m_0$ except through $m_{\mathrm{pole}}$.  We compute $I_\eta$ of Eq.~\eqref{eq:I-eta} with this $\eta(r)$ and with $j_\eta=0$ for $r>r_c$.

In the box of Ref.~\cite{Minamiguchi:2026qcb}, the source-free interval of Eq.~\eqref{eq:source-free} is $r_A=3.99$~fm $\le r\le r_c=13.10$~fm, and Eq.~\eqref{eq:mpole} gives $m_{\mathrm{pole}}=567$~MeV.

\section{Derivation of the \texorpdfstring{$\eta_0$}{eta0} exchange term}\label{app:exchange}

The exchange term is obtained from $\widetilde E_\eta$ of Eq.~\eqref{eq:E-eta} by setting $F\equiv0$ in the quark states and in the vertex, while keeping the profile $\eta(r)$.
The vertex becomes
\begin{equation}
  Y_0^b = \eta(r)\,e^b_r\,i\gamma^0\gamma_5\frac{M}{f_0}\,.
  \label{eq:app-vertex}
\end{equation}
The factor $\Delta_\perp^b/\bm\Delta_\perp^2$ of Eq.~\eqref{eq:Oop} does not depend on the quark states, so Eq.~\eqref{eq:E-eta} reads
\begin{align}
  \widetilde E_\eta^{\mathrm{exch}}(x,\xi,t) &= \frac{M_{\mathrm{cl}}^2}{I\xi}\,\frac{\Delta_\perp^b}{\bm\Delta_\perp^2}\int\!\frac{dz^0}{2\pi}\,e^{iz^0xM_{\mathrm{cl}}}\,R^b(z^0)\,,
  \label{eq:app-def}
\end{align}
where
\begin{align}
  R^b(z^0) &= -M_{\mathrm{cl}}N_c\int\frac{dp^0}{2\pi i}\,e^{-iz^0p^0}\,\mathrm{Tr}\bigl[\gamma^0Y_0^bS\,n_\mu\gamma^\mu\gamma_5
  \nonumber \\
  &\quad\times e^{-iz^0\hat p^3/2}e^{i\bm\Delta\cdot\hat{\bm X}}e^{-iz^0\hat p^3/2}S\bigr]\,,
  \label{eq:app-R}
\end{align}
with $S=i(p^0-H_0)^{-1}\gamma^0$ and $n^\mu=(1,0,0,-1)/M_{\mathrm{cl}}$, where $H_0$ is $H_c$ with $F\equiv0$, and the trace runs over the space, Dirac, and flavor indices.  The residues of the integrand at the occupied states of $H_0$ reproduce the square bracket and the sums over the quark states in Eq.~\eqref{eq:E-eta}.  By the equivalence of the sums over the occupied and the nonoccupied states, the contour is the Feynman contour of $p^0$.

The trace over plane waves, with $q=(0,-\bm\Delta)$ and $z^\mu=z^0(1,0,0,-1)$, for which $q^2=t$ and $n\cdot q=2\xi$, gives
\begin{align}
  R^b(z^0) &= M_{\mathrm{cl}}\,C(z^0;q)\int d^3r\,e^{i\bm\Delta\cdot\bm r}\,\eta(r)\,e^b_r
  \nonumber \\
  &= 4\pi i\,M_{\mathrm{cl}}\,C(z^0;q)\,\frac{\Delta^b}{|\bm\Delta|}\,\eta_1(|\bm\Delta|)\,,
  \label{eq:app-product}
\end{align}
with $\eta_1$ of Eq.~\eqref{eq:eta1}.
The factor $C(z^0;q)$ is the quark loop,
\begin{align}
    C(z^0;q) &= i\frac{2N_c}{f_0}\sum_{i=0}^{2}w_i\mu_i\int\frac{d^4p}{(2\pi)^4}\,e^{-iz\cdot(p-q/2)}
  \nonumber \\
  &\quad\times\mathrm{tr}\left[n_\mu\gamma^\mu\gamma_5S_i(p)\,i\gamma_5S_i(p-q)\right]\,,
  \label{eq:app-loop}
\end{align}
with $S_i(p)=i(\gamma^\nu p_\nu+\mu_i)/(p^2-\mu_i^2)$, where the sum over $i$ is the regularization of Eq.~\eqref{eq:Qreg}.   
With a Feynman parameter $u$, the condition $z^2=0$ leaves the dependence on $z^0$ only in the phase $e^{-i(2u-1)\xi M_{\mathrm{cl}}z^0}$.
The result is
\begin{align}
    C(z^0;q) &= -2i\xi\,f_{\!A}(t)\!\int_0^1\!du\,e^{-i(2u-1)\xi M_{\mathrm{cl}}z^0}\!\phi_\eta(u;t)\,,
  \label{eq:app-C}
\end{align}
with $f_A(t)$ of Eq.~\eqref{eq:GP-exch} and the distribution amplitude
\begin{align}
  \phi_\eta(u;q^2) &= -\frac{1}{J(q^2)}
\sum_{i=0}^{2}w_i\mu_i^2\ln\left[\mu_i^2-u(1-u)q^2\right]\!,
  \label{eq:phi-eta}
\end{align}
where $J(q^2)$ is the regularized vacuum quark loop of Eq.~\eqref{eq:J}.

We insert Eqs.~\eqref{eq:app-product} and \eqref{eq:app-C} into Eq.~\eqref{eq:app-def}.
The contraction $\Delta_\perp^b\Delta^b/\bm\Delta_\perp^2$ equals unity, and the integral over $z^0$ gives $\delta(x-(2u-1)\xi)/M_{\mathrm{cl}}$.
The result is
\begin{align}
  \widetilde E_\eta^{\mathrm{exch}} &= \frac{8\pi M_{\mathrm{cl}}^2}{I}\,f_A(t)\,\frac{\eta_1(|\bm\Delta|)}{|\bm\Delta|}
  \nonumber \\
  &\quad\times\int_0^1du\,\phi_\eta(u;t)\,\delta\big(x-(2u-1)\xi\big)\,.
  \label{eq:app-result}
\end{align}
The delta function fixes $u$ to $(x+\xi)/(2\xi)$ and contributes the Jacobian $1/(2\xi)$.  
This integral over $u$ gives Eqs.~\eqref{eq:E-exch} and \eqref{eq:GP-exch}.

The exchange term of $\widetilde H^{u+d}$ follows from Eqs.~\eqref{eq:Oop-H} and \eqref{eq:H-eta} with $F\equiv0$,
\begin{align}
  \widetilde H_\eta^{\mathrm{exch}}(x,\xi,t) &= \frac{M_{\mathrm{cl}}}{2I}\left(\delta^{b3}+2M_{\mathrm{cl}}\xi\frac{\Delta_\perp^b}{\bm\Delta_\perp^2}\right)
  \nonumber \\
  &\quad\times\int\frac{dz^0}{2\pi}\,e^{iz^0xM_{\mathrm{cl}}}\,R^b(z^0)\,.
\end{align}
Equation~\eqref{eq:app-product} shows that $R^b(z^0)$ is proportional to $\Delta^b$.  The contraction with the factor in the parentheses is $\Delta^3+2M_{\mathrm{cl}}\xi$, which vanishes by Eq.~\eqref{eq:kinematics}.  This result is Eq.~\eqref{eq:H-exch}.

\bibliography{gpd}

\begin{thebibliography}{66}%
\makeatletter
\providecommand \@ifxundefined [1]{%
 \@ifx{#1\undefined}
}%
\providecommand \@ifnum [1]{%
 \ifnum #1\expandafter \@firstoftwo
 \else \expandafter \@secondoftwo
 \fi
}%
\providecommand \@ifx [1]{%
 \ifx #1\expandafter \@firstoftwo
 \else \expandafter \@secondoftwo
 \fi
}%
\providecommand \natexlab [1]{#1}%
\providecommand \enquote  [1]{``#1''}%
\providecommand \bibnamefont  [1]{#1}%
\providecommand \bibfnamefont [1]{#1}%
\providecommand \citenamefont [1]{#1}%
\providecommand \href@noop [0]{\@secondoftwo}%
\providecommand \href [0]{\begingroup \@sanitize@url \@href}%
\providecommand \@href[1]{\@@startlink{#1}\@@href}%
\providecommand \@@href[1]{\endgroup#1\@@endlink}%
\providecommand \@sanitize@url [0]{\catcode `\\12\catcode `\$12\catcode `\&12\catcode `\#12\catcode `\^12\catcode `\_12\catcode `\%12\relax}%
\providecommand \@@startlink[1]{}%
\providecommand \@@endlink[0]{}%
\providecommand \url  [0]{\begingroup\@sanitize@url \@url }%
\providecommand \@url [1]{\endgroup\@href {#1}{\urlprefix }}%
\providecommand \urlprefix  [0]{URL }%
\providecommand \Eprint [0]{\href }%
\providecommand \doibase [0]{https://doi.org/}%
\providecommand \selectlanguage [0]{\@gobble}%
\providecommand \bibinfo  [0]{\@secondoftwo}%
\providecommand \bibfield  [0]{\@secondoftwo}%
\providecommand \translation [1]{[#1]}%
\providecommand \BibitemOpen [0]{}%
\providecommand \bibitemStop [0]{}%
\providecommand \bibitemNoStop [0]{.\EOS\space}%
\providecommand \EOS [0]{\spacefactor3000\relax}%
\providecommand \BibitemShut  [1]{\csname bibitem#1\endcsname}%
\let\auto@bib@innerbib\@empty
\bibitem [{\citenamefont {Adler}(1969)}]{Adler:1969gk}%
  \BibitemOpen
  \bibfield  {author} {\bibinfo {author} {\bibfnamefont {S.~L.}\ \bibnamefont {Adler}},\ }\href {https://doi.org/10.1103/PhysRev.177.2426} {\bibfield  {journal} {\bibinfo  {journal} {Phys. Rev.}\ }\textbf {\bibinfo {volume} {177}},\ \bibinfo {pages} {2426} (\bibinfo {year} {1969})}\BibitemShut {NoStop}%
\bibitem [{\citenamefont {Bell}\ and\ \citenamefont {Jackiw}(1969)}]{Bell:1969ts}%
  \BibitemOpen
  \bibfield  {author} {\bibinfo {author} {\bibfnamefont {J.~S.}\ \bibnamefont {Bell}}\ and\ \bibinfo {author} {\bibfnamefont {R.}~\bibnamefont {Jackiw}},\ }\href {https://doi.org/10.1007/BF02823296} {\bibfield  {journal} {\bibinfo  {journal} {Nuovo Cim. A}\ }\textbf {\bibinfo {volume} {60}},\ \bibinfo {pages} {47} (\bibinfo {year} {1969})}\BibitemShut {NoStop}%
\bibitem [{\citenamefont {Kodaira}(1980)}]{Kodaira:1979pa}%
  \BibitemOpen
  \bibfield  {author} {\bibinfo {author} {\bibfnamefont {J.}~\bibnamefont {Kodaira}},\ }\href {https://doi.org/10.1016/0550-3213(80)90310-7} {\bibfield  {journal} {\bibinfo  {journal} {Nucl. Phys. B}\ }\textbf {\bibinfo {volume} {165}},\ \bibinfo {pages} {129} (\bibinfo {year} {1980})}\BibitemShut {NoStop}%
\bibitem [{\citenamefont {Jaffe}\ and\ \citenamefont {Manohar}(1990)}]{Jaffe:1989jz}%
  \BibitemOpen
  \bibfield  {author} {\bibinfo {author} {\bibfnamefont {R.~L.}\ \bibnamefont {Jaffe}}\ and\ \bibinfo {author} {\bibfnamefont {A.}~\bibnamefont {Manohar}},\ }\href {https://doi.org/10.1016/0550-3213(90)90506-9} {\bibfield  {journal} {\bibinfo  {journal} {Nucl. Phys. B}\ }\textbf {\bibinfo {volume} {337}},\ \bibinfo {pages} {509} (\bibinfo {year} {1990})}\BibitemShut {NoStop}%
\bibitem [{\citenamefont {Ashman}\ \emph {et~al.}(1988)\citenamefont {Ashman} \emph {et~al.}}]{EuropeanMuon:1987isl}%
  \BibitemOpen
  \bibfield  {author} {\bibinfo {author} {\bibfnamefont {J.}~\bibnamefont {Ashman}} \emph {et~al.} (\bibinfo {collaboration} {European Muon}),\ }\href {https://doi.org/10.1016/0370-2693(88)91523-7} {\bibfield  {journal} {\bibinfo  {journal} {Phys. Lett. B}\ }\textbf {\bibinfo {volume} {206}},\ \bibinfo {pages} {364} (\bibinfo {year} {1988})}\BibitemShut {NoStop}%
\bibitem [{\citenamefont {Altarelli}\ and\ \citenamefont {Ross}(1988)}]{Altarelli:1988nr}%
  \BibitemOpen
  \bibfield  {author} {\bibinfo {author} {\bibfnamefont {G.}~\bibnamefont {Altarelli}}\ and\ \bibinfo {author} {\bibfnamefont {G.~G.}\ \bibnamefont {Ross}},\ }\href {https://doi.org/10.1016/0370-2693(88)91335-4} {\bibfield  {journal} {\bibinfo  {journal} {Phys. Lett. B}\ }\textbf {\bibinfo {volume} {212}},\ \bibinfo {pages} {391} (\bibinfo {year} {1988})}\BibitemShut {NoStop}%
\bibitem [{\citenamefont {Carlitz}\ \emph {et~al.}(1988)\citenamefont {Carlitz}, \citenamefont {Collins},\ and\ \citenamefont {Mueller}}]{Carlitz:1988ab}%
  \BibitemOpen
  \bibfield  {author} {\bibinfo {author} {\bibfnamefont {R.~D.}\ \bibnamefont {Carlitz}}, \bibinfo {author} {\bibfnamefont {J.~C.}\ \bibnamefont {Collins}},\ and\ \bibinfo {author} {\bibfnamefont {A.~H.}\ \bibnamefont {Mueller}},\ }\href {https://doi.org/10.1016/0370-2693(88)91474-8} {\bibfield  {journal} {\bibinfo  {journal} {Phys. Lett. B}\ }\textbf {\bibinfo {volume} {214}},\ \bibinfo {pages} {229} (\bibinfo {year} {1988})}\BibitemShut {NoStop}%
\bibitem [{\citenamefont {Veneziano}(1989)}]{Veneziano:1989ei}%
  \BibitemOpen
  \bibfield  {author} {\bibinfo {author} {\bibfnamefont {G.}~\bibnamefont {Veneziano}},\ }\href {https://doi.org/10.1142/S0217732389001830} {\bibfield  {journal} {\bibinfo  {journal} {Mod. Phys. Lett. A}\ }\textbf {\bibinfo {volume} {4}},\ \bibinfo {pages} {1605} (\bibinfo {year} {1989})}\BibitemShut {NoStop}%
\bibitem [{\citenamefont {Narison}\ \emph {et~al.}(1995)\citenamefont {Narison}, \citenamefont {Shore},\ and\ \citenamefont {Veneziano}}]{Narison:1994hv}%
  \BibitemOpen
  \bibfield  {author} {\bibinfo {author} {\bibfnamefont {S.}~\bibnamefont {Narison}}, \bibinfo {author} {\bibfnamefont {G.~M.}\ \bibnamefont {Shore}},\ and\ \bibinfo {author} {\bibfnamefont {G.}~\bibnamefont {Veneziano}},\ }\href {https://doi.org/10.1016/0550-3213(94)00329-D} {\bibfield  {journal} {\bibinfo  {journal} {Nucl. Phys. B}\ }\textbf {\bibinfo {volume} {433}},\ \bibinfo {pages} {209} (\bibinfo {year} {1995})},\ \Eprint {https://arxiv.org/abs/hep-ph/9404277} {arXiv:hep-ph/9404277} \BibitemShut {NoStop}%
\bibitem [{\citenamefont {Hatta}(2020)}]{Hatta:2020ltd}%
  \BibitemOpen
  \bibfield  {author} {\bibinfo {author} {\bibfnamefont {Y.}~\bibnamefont {Hatta}},\ }\href {https://doi.org/10.1103/PhysRevD.102.094004} {\bibfield  {journal} {\bibinfo  {journal} {Phys. Rev. D}\ }\textbf {\bibinfo {volume} {102}},\ \bibinfo {pages} {094004} (\bibinfo {year} {2020})},\ \Eprint {https://arxiv.org/abs/2009.03657} {arXiv:2009.03657 [hep-ph]} \BibitemShut {NoStop}%
\bibitem [{\citenamefont {Tarasov}\ and\ \citenamefont {Venugopalan}(2020)}]{Tarasov:2020cwl}%
  \BibitemOpen
  \bibfield  {author} {\bibinfo {author} {\bibfnamefont {A.}~\bibnamefont {Tarasov}}\ and\ \bibinfo {author} {\bibfnamefont {R.}~\bibnamefont {Venugopalan}},\ }\href {https://doi.org/10.1103/PhysRevD.102.114022} {\bibfield  {journal} {\bibinfo  {journal} {Phys. Rev. D}\ }\textbf {\bibinfo {volume} {102}},\ \bibinfo {pages} {114022} (\bibinfo {year} {2020})},\ \Eprint {https://arxiv.org/abs/2008.08104} {arXiv:2008.08104 [hep-ph]} \BibitemShut {NoStop}%
\bibitem [{\citenamefont {Tarasov}\ and\ \citenamefont {Venugopalan}(2022)}]{Tarasov:2021yll}%
  \BibitemOpen
  \bibfield  {author} {\bibinfo {author} {\bibfnamefont {A.}~\bibnamefont {Tarasov}}\ and\ \bibinfo {author} {\bibfnamefont {R.}~\bibnamefont {Venugopalan}},\ }\href {https://doi.org/10.1103/PhysRevD.105.014020} {\bibfield  {journal} {\bibinfo  {journal} {Phys. Rev. D}\ }\textbf {\bibinfo {volume} {105}},\ \bibinfo {pages} {014020} (\bibinfo {year} {2022})},\ \Eprint {https://arxiv.org/abs/2109.10370} {arXiv:2109.10370 [hep-ph]} \BibitemShut {NoStop}%
\bibitem [{\citenamefont {Bhattacharya}\ \emph {et~al.}(2023{\natexlab{a}})\citenamefont {Bhattacharya}, \citenamefont {Hatta},\ and\ \citenamefont {Vogelsang}}]{Bhattacharya:2022xxw}%
  \BibitemOpen
  \bibfield  {author} {\bibinfo {author} {\bibfnamefont {S.}~\bibnamefont {Bhattacharya}}, \bibinfo {author} {\bibfnamefont {Y.}~\bibnamefont {Hatta}},\ and\ \bibinfo {author} {\bibfnamefont {W.}~\bibnamefont {Vogelsang}},\ }\href {https://doi.org/10.1103/PhysRevD.107.014026} {\bibfield  {journal} {\bibinfo  {journal} {Phys. Rev. D}\ }\textbf {\bibinfo {volume} {107}},\ \bibinfo {pages} {014026} (\bibinfo {year} {2023}{\natexlab{a}})},\ \Eprint {https://arxiv.org/abs/2210.13419} {arXiv:2210.13419 [hep-ph]} \BibitemShut {NoStop}%
\bibitem [{\citenamefont {Bhattacharya}\ \emph {et~al.}(2023{\natexlab{b}})\citenamefont {Bhattacharya}, \citenamefont {Hatta},\ and\ \citenamefont {Vogelsang}}]{Bhattacharya:2023wvy}%
  \BibitemOpen
  \bibfield  {author} {\bibinfo {author} {\bibfnamefont {S.}~\bibnamefont {Bhattacharya}}, \bibinfo {author} {\bibfnamefont {Y.}~\bibnamefont {Hatta}},\ and\ \bibinfo {author} {\bibfnamefont {W.}~\bibnamefont {Vogelsang}},\ }\href {https://doi.org/10.1103/PhysRevD.108.014029} {\bibfield  {journal} {\bibinfo  {journal} {Phys. Rev. D}\ }\textbf {\bibinfo {volume} {108}},\ \bibinfo {pages} {014029} (\bibinfo {year} {2023}{\natexlab{b}})},\ \Eprint {https://arxiv.org/abs/2305.09431} {arXiv:2305.09431 [hep-ph]} \BibitemShut {NoStop}%
\bibitem [{\citenamefont {Bhattacharya}\ \emph {et~al.}(2025)\citenamefont {Bhattacharya}, \citenamefont {Hatta},\ and\ \citenamefont {Schoenleber}}]{Bhattacharya:2024geo}%
  \BibitemOpen
  \bibfield  {author} {\bibinfo {author} {\bibfnamefont {S.}~\bibnamefont {Bhattacharya}}, \bibinfo {author} {\bibfnamefont {Y.}~\bibnamefont {Hatta}},\ and\ \bibinfo {author} {\bibfnamefont {J.}~\bibnamefont {Schoenleber}},\ }\href {https://doi.org/10.1103/PhysRevD.111.014013} {\bibfield  {journal} {\bibinfo  {journal} {Phys. Rev. D}\ }\textbf {\bibinfo {volume} {111}},\ \bibinfo {pages} {014013} (\bibinfo {year} {2025})},\ \Eprint {https://arxiv.org/abs/2411.07024} {arXiv:2411.07024 [hep-ph]} \BibitemShut {NoStop}%
\bibitem [{\citenamefont {Tarasov}\ and\ \citenamefont {Venugopalan}(2025)}]{Tarasov:2025mvn}%
  \BibitemOpen
  \bibfield  {author} {\bibinfo {author} {\bibfnamefont {A.}~\bibnamefont {Tarasov}}\ and\ \bibinfo {author} {\bibfnamefont {R.}~\bibnamefont {Venugopalan}},\ }\href {https://doi.org/10.1103/PhysRevD.111.074027} {\bibfield  {journal} {\bibinfo  {journal} {Phys. Rev. D}\ }\textbf {\bibinfo {volume} {111}},\ \bibinfo {pages} {074027} (\bibinfo {year} {2025})},\ \Eprint {https://arxiv.org/abs/2501.10519} {arXiv:2501.10519 [hep-ph]} \BibitemShut {NoStop}%
\bibitem [{\citenamefont {Fukushima}\ and\ \citenamefont {Uji}(2026)}]{Fukushima:2026sot}%
  \BibitemOpen
  \bibfield  {author} {\bibinfo {author} {\bibfnamefont {K.}~\bibnamefont {Fukushima}}\ and\ \bibinfo {author} {\bibfnamefont {T.}~\bibnamefont {Uji}},\ }\href@noop {} {\  (\bibinfo {year} {2026})},\ \Eprint {https://arxiv.org/abs/2607.03123} {arXiv:2607.03123 [hep-ph]} \BibitemShut {NoStop}%
\bibitem [{\citenamefont {Minamiguchi}\ and\ \citenamefont {Uji}(2026{\natexlab{a}})}]{Minamiguchi:2026uaq}%
  \BibitemOpen
  \bibfield  {author} {\bibinfo {author} {\bibfnamefont {J.}~\bibnamefont {Minamiguchi}}\ and\ \bibinfo {author} {\bibfnamefont {T.}~\bibnamefont {Uji}},\ }\href@noop {} {\  (\bibinfo {year} {2026}{\natexlab{a}})},\ \Eprint {https://arxiv.org/abs/2609.25867} {arXiv:2609.25867 [hep-ph]} \BibitemShut {NoStop}%
\bibitem [{\citenamefont {Shore}\ and\ \citenamefont {Veneziano}(1992)}]{Shore:1991dv}%
  \BibitemOpen
  \bibfield  {author} {\bibinfo {author} {\bibfnamefont {G.~M.}\ \bibnamefont {Shore}}\ and\ \bibinfo {author} {\bibfnamefont {G.}~\bibnamefont {Veneziano}},\ }\href {https://doi.org/10.1016/0550-3213(92)90639-S} {\bibfield  {journal} {\bibinfo  {journal} {Nucl. Phys. B}\ }\textbf {\bibinfo {volume} {381}},\ \bibinfo {pages} {23} (\bibinfo {year} {1992})}\BibitemShut {NoStop}%
\bibitem [{\citenamefont {Ji}(1997)}]{Ji:1996ek}%
  \BibitemOpen
  \bibfield  {author} {\bibinfo {author} {\bibfnamefont {X.-D.}\ \bibnamefont {Ji}},\ }\href {https://doi.org/10.1103/PhysRevLett.78.610} {\bibfield  {journal} {\bibinfo  {journal} {Phys. Rev. Lett.}\ }\textbf {\bibinfo {volume} {78}},\ \bibinfo {pages} {610} (\bibinfo {year} {1997})},\ \Eprint {https://arxiv.org/abs/hep-ph/9603249} {arXiv:hep-ph/9603249} \BibitemShut {NoStop}%
\bibitem [{\citenamefont {Witten}(1979)}]{Witten:1979vv}%
  \BibitemOpen
  \bibfield  {author} {\bibinfo {author} {\bibfnamefont {E.}~\bibnamefont {Witten}},\ }\href {https://doi.org/10.1016/0550-3213(79)90031-2} {\bibfield  {journal} {\bibinfo  {journal} {Nucl. Phys. B}\ }\textbf {\bibinfo {volume} {156}},\ \bibinfo {pages} {269} (\bibinfo {year} {1979})}\BibitemShut {NoStop}%
\bibitem [{\citenamefont {Veneziano}(1979)}]{Veneziano:1979ec}%
  \BibitemOpen
  \bibfield  {author} {\bibinfo {author} {\bibfnamefont {G.}~\bibnamefont {Veneziano}},\ }\href {https://doi.org/10.1016/0550-3213(79)90332-8} {\bibfield  {journal} {\bibinfo  {journal} {Nucl. Phys. B}\ }\textbf {\bibinfo {volume} {159}},\ \bibinfo {pages} {213} (\bibinfo {year} {1979})}\BibitemShut {NoStop}%
\bibitem [{\citenamefont {Fujii}\ \emph {et~al.}(2025)\citenamefont {Fujii}, \citenamefont {Kawaguchi},\ and\ \citenamefont {Tanaka}}]{Fujii:2025aip}%
  \BibitemOpen
  \bibfield  {author} {\bibinfo {author} {\bibfnamefont {D.}~\bibnamefont {Fujii}}, \bibinfo {author} {\bibfnamefont {M.}~\bibnamefont {Kawaguchi}},\ and\ \bibinfo {author} {\bibfnamefont {M.}~\bibnamefont {Tanaka}},\ }\href {https://doi.org/10.1016/j.physletb.2025.139559} {\bibfield  {journal} {\bibinfo  {journal} {Phys. Lett. B}\ }\textbf {\bibinfo {volume} {866}},\ \bibinfo {pages} {139559} (\bibinfo {year} {2025})},\ \Eprint {https://arxiv.org/abs/2503.09686} {arXiv:2503.09686 [hep-ph]} \BibitemShut {NoStop}%
\bibitem [{\citenamefont {Tanaka}\ \emph {et~al.}(2025)\citenamefont {Tanaka}, \citenamefont {Fujii},\ and\ \citenamefont {Kawaguchi}}]{Tanaka:2025pny}%
  \BibitemOpen
  \bibfield  {author} {\bibinfo {author} {\bibfnamefont {M.}~\bibnamefont {Tanaka}}, \bibinfo {author} {\bibfnamefont {D.}~\bibnamefont {Fujii}},\ and\ \bibinfo {author} {\bibfnamefont {M.}~\bibnamefont {Kawaguchi}},\ }\href {https://doi.org/10.1103/gzj5-7bln} {\bibfield  {journal} {\bibinfo  {journal} {Phys. Rev. D}\ }\textbf {\bibinfo {volume} {112}},\ \bibinfo {pages} {054048} (\bibinfo {year} {2025})},\ \Eprint {https://arxiv.org/abs/2507.21220} {arXiv:2507.21220 [hep-ph]} \BibitemShut {NoStop}%
\bibitem [{\citenamefont {Kawaguchi}\ \emph {et~al.}(2026)\citenamefont {Kawaguchi}, \citenamefont {Tanaka},\ and\ \citenamefont {Tanaka}}]{Kawaguchi:2026yae}%
  \BibitemOpen
  \bibfield  {author} {\bibinfo {author} {\bibfnamefont {M.}~\bibnamefont {Kawaguchi}}, \bibinfo {author} {\bibfnamefont {K.}~\bibnamefont {Tanaka}},\ and\ \bibinfo {author} {\bibfnamefont {M.}~\bibnamefont {Tanaka}},\ }\href@noop {} {\  (\bibinfo {year} {2026})},\ \Eprint {https://arxiv.org/abs/2608.30434} {arXiv:2608.30434 [hep-ph]} \BibitemShut {NoStop}%
\bibitem [{\citenamefont {Diakonov}\ \emph {et~al.}(1988)\citenamefont {Diakonov}, \citenamefont {Petrov},\ and\ \citenamefont {Pobylitsa}}]{Diakonov:1987ty}%
  \BibitemOpen
  \bibfield  {author} {\bibinfo {author} {\bibfnamefont {D.}~\bibnamefont {Diakonov}}, \bibinfo {author} {\bibfnamefont {V.~Y.}\ \bibnamefont {Petrov}},\ and\ \bibinfo {author} {\bibfnamefont {P.~V.}\ \bibnamefont {Pobylitsa}},\ }\href {https://doi.org/10.1016/0550-3213(88)90443-9} {\bibfield  {journal} {\bibinfo  {journal} {Nucl. Phys. B}\ }\textbf {\bibinfo {volume} {306}},\ \bibinfo {pages} {809} (\bibinfo {year} {1988})}\BibitemShut {NoStop}%
\bibitem [{\citenamefont {Reinhardt}\ and\ \citenamefont {Wunsch}(1988)}]{Reinhardt:1988fz}%
  \BibitemOpen
  \bibfield  {author} {\bibinfo {author} {\bibfnamefont {H.}~\bibnamefont {Reinhardt}}\ and\ \bibinfo {author} {\bibfnamefont {R.}~\bibnamefont {Wunsch}},\ }\href {https://doi.org/10.1016/0370-2693(88)91363-9} {\bibfield  {journal} {\bibinfo  {journal} {Phys. Lett. B}\ }\textbf {\bibinfo {volume} {215}},\ \bibinfo {pages} {577} (\bibinfo {year} {1988})}\BibitemShut {NoStop}%
\bibitem [{\citenamefont {Alkofer}\ \emph {et~al.}(1996)\citenamefont {Alkofer}, \citenamefont {Reinhardt},\ and\ \citenamefont {Weigel}}]{Alkofer:1994ph}%
  \BibitemOpen
  \bibfield  {author} {\bibinfo {author} {\bibfnamefont {R.}~\bibnamefont {Alkofer}}, \bibinfo {author} {\bibfnamefont {H.}~\bibnamefont {Reinhardt}},\ and\ \bibinfo {author} {\bibfnamefont {H.}~\bibnamefont {Weigel}},\ }\href {https://doi.org/10.1016/0370-1573(95)00018-6} {\bibfield  {journal} {\bibinfo  {journal} {Phys. Rept.}\ }\textbf {\bibinfo {volume} {265}},\ \bibinfo {pages} {139} (\bibinfo {year} {1996})},\ \Eprint {https://arxiv.org/abs/hep-ph/9501213} {arXiv:hep-ph/9501213} \BibitemShut {NoStop}%
\bibitem [{\citenamefont {Christov}\ \emph {et~al.}(1996)\citenamefont {Christov}, \citenamefont {Blotz}, \citenamefont {Kim}, \citenamefont {Pobylitsa}, \citenamefont {Watabe}, \citenamefont {Meissner}, \citenamefont {Ruiz~Arriola},\ and\ \citenamefont {Goeke}}]{Christov:1995vm}%
  \BibitemOpen
  \bibfield  {author} {\bibinfo {author} {\bibfnamefont {C.~V.}\ \bibnamefont {Christov}}, \bibinfo {author} {\bibfnamefont {A.}~\bibnamefont {Blotz}}, \bibinfo {author} {\bibfnamefont {H.-C.}\ \bibnamefont {Kim}}, \bibinfo {author} {\bibfnamefont {P.}~\bibnamefont {Pobylitsa}}, \bibinfo {author} {\bibfnamefont {T.}~\bibnamefont {Watabe}}, \bibinfo {author} {\bibfnamefont {T.}~\bibnamefont {Meissner}}, \bibinfo {author} {\bibfnamefont {E.}~\bibnamefont {Ruiz~Arriola}},\ and\ \bibinfo {author} {\bibfnamefont {K.}~\bibnamefont {Goeke}},\ }\href {https://doi.org/10.1016/0146-6410(96)00057-9} {\bibfield  {journal} {\bibinfo  {journal} {Prog. Part. Nucl. Phys.}\ }\textbf {\bibinfo {volume} {37}},\ \bibinfo {pages} {91} (\bibinfo {year} {1996})},\ \Eprint {https://arxiv.org/abs/hep-ph/9604441} {arXiv:hep-ph/9604441} \BibitemShut {NoStop}%
\bibitem [{\citenamefont {Diakonov}(1997)}]{Diakonov:1997sj}%
  \BibitemOpen
  \bibfield  {author} {\bibinfo {author} {\bibfnamefont {D.}~\bibnamefont {Diakonov}},\ }in\ \href@noop {} {\emph {\bibinfo {booktitle} {{Advanced Summer School on Nonperturbative Quantum Field Physics}}}}\ (\bibinfo {year} {1997})\ pp.\ \bibinfo {pages} {1--55},\ \Eprint {https://arxiv.org/abs/hep-ph/9802298} {arXiv:hep-ph/9802298} \BibitemShut {NoStop}%
\bibitem [{\citenamefont {Weigel}\ \emph {et~al.}(1997{\natexlab{a}})\citenamefont {Weigel}, \citenamefont {Gamberg},\ and\ \citenamefont {Reinhardt}}]{Weigel:1996kw}%
  \BibitemOpen
  \bibfield  {author} {\bibinfo {author} {\bibfnamefont {H.}~\bibnamefont {Weigel}}, \bibinfo {author} {\bibfnamefont {L.~P.}\ \bibnamefont {Gamberg}},\ and\ \bibinfo {author} {\bibfnamefont {H.}~\bibnamefont {Reinhardt}},\ }\href {https://doi.org/10.1016/S0370-2693(97)00296-7} {\bibfield  {journal} {\bibinfo  {journal} {Phys. Lett. B}\ }\textbf {\bibinfo {volume} {399}},\ \bibinfo {pages} {287} (\bibinfo {year} {1997}{\natexlab{a}})},\ \Eprint {https://arxiv.org/abs/hep-ph/9604295} {arXiv:hep-ph/9604295} \BibitemShut {NoStop}%
\bibitem [{\citenamefont {Diakonov}\ \emph {et~al.}(1996)\citenamefont {Diakonov}, \citenamefont {Petrov}, \citenamefont {Pobylitsa}, \citenamefont {Polyakov},\ and\ \citenamefont {Weiss}}]{Diakonov:1996sr}%
  \BibitemOpen
  \bibfield  {author} {\bibinfo {author} {\bibfnamefont {D.}~\bibnamefont {Diakonov}}, \bibinfo {author} {\bibfnamefont {V.}~\bibnamefont {Petrov}}, \bibinfo {author} {\bibfnamefont {P.}~\bibnamefont {Pobylitsa}}, \bibinfo {author} {\bibfnamefont {M.~V.}\ \bibnamefont {Polyakov}},\ and\ \bibinfo {author} {\bibfnamefont {C.}~\bibnamefont {Weiss}},\ }\href {https://doi.org/10.1016/S0550-3213(96)00486-5} {\bibfield  {journal} {\bibinfo  {journal} {Nucl. Phys. B}\ }\textbf {\bibinfo {volume} {480}},\ \bibinfo {pages} {341} (\bibinfo {year} {1996})},\ \Eprint {https://arxiv.org/abs/hep-ph/9606314} {arXiv:hep-ph/9606314} \BibitemShut {NoStop}%
\bibitem [{\citenamefont {Weigel}\ \emph {et~al.}(1997{\natexlab{b}})\citenamefont {Weigel}, \citenamefont {Gamberg},\ and\ \citenamefont {Reinhardt}}]{Weigel:1996jh}%
  \BibitemOpen
  \bibfield  {author} {\bibinfo {author} {\bibfnamefont {H.}~\bibnamefont {Weigel}}, \bibinfo {author} {\bibfnamefont {L.~P.}\ \bibnamefont {Gamberg}},\ and\ \bibinfo {author} {\bibfnamefont {H.}~\bibnamefont {Reinhardt}},\ }\href {https://doi.org/10.1103/PhysRevD.55.6910} {\bibfield  {journal} {\bibinfo  {journal} {Phys. Rev. D}\ }\textbf {\bibinfo {volume} {55}},\ \bibinfo {pages} {6910} (\bibinfo {year} {1997}{\natexlab{b}})},\ \Eprint {https://arxiv.org/abs/hep-ph/9609226} {arXiv:hep-ph/9609226} \BibitemShut {NoStop}%
\bibitem [{\citenamefont {Diakonov}\ \emph {et~al.}(1997)\citenamefont {Diakonov}, \citenamefont {Petrov}, \citenamefont {Pobylitsa}, \citenamefont {Polyakov},\ and\ \citenamefont {Weiss}}]{Diakonov:1997vc}%
  \BibitemOpen
  \bibfield  {author} {\bibinfo {author} {\bibfnamefont {D.}~\bibnamefont {Diakonov}}, \bibinfo {author} {\bibfnamefont {V.~Y.}\ \bibnamefont {Petrov}}, \bibinfo {author} {\bibfnamefont {P.~V.}\ \bibnamefont {Pobylitsa}}, \bibinfo {author} {\bibfnamefont {M.~V.}\ \bibnamefont {Polyakov}},\ and\ \bibinfo {author} {\bibfnamefont {C.}~\bibnamefont {Weiss}},\ }\href {https://doi.org/10.1103/PhysRevD.56.4069} {\bibfield  {journal} {\bibinfo  {journal} {Phys. Rev. D}\ }\textbf {\bibinfo {volume} {56}},\ \bibinfo {pages} {4069} (\bibinfo {year} {1997})},\ \Eprint {https://arxiv.org/abs/hep-ph/9703420} {arXiv:hep-ph/9703420} \BibitemShut {NoStop}%
\bibitem [{\citenamefont {Weiss}\ and\ \citenamefont {Goeke}(1997)}]{Weiss:1997rt}%
  \BibitemOpen
  \bibfield  {author} {\bibinfo {author} {\bibfnamefont {C.}~\bibnamefont {Weiss}}\ and\ \bibinfo {author} {\bibfnamefont {K.}~\bibnamefont {Goeke}},\ }\href@noop {} {\  (\bibinfo {year} {1997})},\ \Eprint {https://arxiv.org/abs/hep-ph/9712447} {arXiv:hep-ph/9712447} \BibitemShut {NoStop}%
\bibitem [{\citenamefont {Wakamatsu}\ and\ \citenamefont {Kubota}(1998)}]{Wakamatsu:1997en}%
  \BibitemOpen
  \bibfield  {author} {\bibinfo {author} {\bibfnamefont {M.}~\bibnamefont {Wakamatsu}}\ and\ \bibinfo {author} {\bibfnamefont {T.}~\bibnamefont {Kubota}},\ }\href {https://doi.org/10.1103/PhysRevD.57.5755} {\bibfield  {journal} {\bibinfo  {journal} {Phys. Rev. D}\ }\textbf {\bibinfo {volume} {57}},\ \bibinfo {pages} {5755} (\bibinfo {year} {1998})},\ \Eprint {https://arxiv.org/abs/hep-ph/9707500} {arXiv:hep-ph/9707500} \BibitemShut {NoStop}%
\bibitem [{\citenamefont {Pobylitsa}\ \emph {et~al.}(1999)\citenamefont {Pobylitsa}, \citenamefont {Polyakov}, \citenamefont {Goeke}, \citenamefont {Watabe},\ and\ \citenamefont {Weiss}}]{Pobylitsa:1998tk}%
  \BibitemOpen
  \bibfield  {author} {\bibinfo {author} {\bibfnamefont {P.~V.}\ \bibnamefont {Pobylitsa}}, \bibinfo {author} {\bibfnamefont {M.~V.}\ \bibnamefont {Polyakov}}, \bibinfo {author} {\bibfnamefont {K.}~\bibnamefont {Goeke}}, \bibinfo {author} {\bibfnamefont {T.}~\bibnamefont {Watabe}},\ and\ \bibinfo {author} {\bibfnamefont {C.}~\bibnamefont {Weiss}},\ }\href {https://doi.org/10.1103/PhysRevD.59.034024} {\bibfield  {journal} {\bibinfo  {journal} {Phys. Rev. D}\ }\textbf {\bibinfo {volume} {59}},\ \bibinfo {pages} {034024} (\bibinfo {year} {1999})},\ \Eprint {https://arxiv.org/abs/hep-ph/9804436} {arXiv:hep-ph/9804436} \BibitemShut {NoStop}%
\bibitem [{\citenamefont {Wakamatsu}\ and\ \citenamefont {Kubota}(1999)}]{Wakamatsu:1998rx}%
  \BibitemOpen
  \bibfield  {author} {\bibinfo {author} {\bibfnamefont {M.}~\bibnamefont {Wakamatsu}}\ and\ \bibinfo {author} {\bibfnamefont {T.}~\bibnamefont {Kubota}},\ }\href {https://doi.org/10.1103/PhysRevD.60.034020} {\bibfield  {journal} {\bibinfo  {journal} {Phys. Rev. D}\ }\textbf {\bibinfo {volume} {60}},\ \bibinfo {pages} {034020} (\bibinfo {year} {1999})},\ \Eprint {https://arxiv.org/abs/hep-ph/9809443} {arXiv:hep-ph/9809443} \BibitemShut {NoStop}%
\bibitem [{\citenamefont {Weigel}\ \emph {et~al.}(1999)\citenamefont {Weigel}, \citenamefont {Ruiz~Arriola},\ and\ \citenamefont {Gamberg}}]{Weigel:1999pc}%
  \BibitemOpen
  \bibfield  {author} {\bibinfo {author} {\bibfnamefont {H.}~\bibnamefont {Weigel}}, \bibinfo {author} {\bibfnamefont {E.}~\bibnamefont {Ruiz~Arriola}},\ and\ \bibinfo {author} {\bibfnamefont {L.~P.}\ \bibnamefont {Gamberg}},\ }\href {https://doi.org/10.1016/S0550-3213(99)00426-5} {\bibfield  {journal} {\bibinfo  {journal} {Nucl. Phys. B}\ }\textbf {\bibinfo {volume} {560}},\ \bibinfo {pages} {383} (\bibinfo {year} {1999})},\ \Eprint {https://arxiv.org/abs/hep-ph/9905329} {arXiv:hep-ph/9905329} \BibitemShut {NoStop}%
\bibitem [{\citenamefont {Wakamatsu}\ and\ \citenamefont {Watabe}(2000)}]{Wakamatsu:1999nj}%
  \BibitemOpen
  \bibfield  {author} {\bibinfo {author} {\bibfnamefont {M.}~\bibnamefont {Wakamatsu}}\ and\ \bibinfo {author} {\bibfnamefont {T.}~\bibnamefont {Watabe}},\ }\href {https://doi.org/10.1103/PhysRevD.62.054009} {\bibfield  {journal} {\bibinfo  {journal} {Phys. Rev. D}\ }\textbf {\bibinfo {volume} {62}},\ \bibinfo {pages} {054009} (\bibinfo {year} {2000})},\ \Eprint {https://arxiv.org/abs/hep-ph/9912500} {arXiv:hep-ph/9912500} \BibitemShut {NoStop}%
\bibitem [{\citenamefont {Takyi}\ and\ \citenamefont {Weigel}(2019)}]{Takyi:2019ahv}%
  \BibitemOpen
  \bibfield  {author} {\bibinfo {author} {\bibfnamefont {I.}~\bibnamefont {Takyi}}\ and\ \bibinfo {author} {\bibfnamefont {H.}~\bibnamefont {Weigel}},\ }\href {https://doi.org/10.1140/epja/i2019-12806-3} {\bibfield  {journal} {\bibinfo  {journal} {Eur. Phys. J. A}\ }\textbf {\bibinfo {volume} {55}},\ \bibinfo {pages} {128} (\bibinfo {year} {2019})},\ \Eprint {https://arxiv.org/abs/1903.11435} {arXiv:1903.11435 [hep-ph]} \BibitemShut {NoStop}%
\bibitem [{\citenamefont {Petrov}\ \emph {et~al.}(1998)\citenamefont {Petrov}, \citenamefont {Pobylitsa}, \citenamefont {Polyakov}, \citenamefont {Bornig}, \citenamefont {Goeke},\ and\ \citenamefont {Weiss}}]{Petrov:1998kf}%
  \BibitemOpen
  \bibfield  {author} {\bibinfo {author} {\bibfnamefont {V.~Y.}\ \bibnamefont {Petrov}}, \bibinfo {author} {\bibfnamefont {P.~V.}\ \bibnamefont {Pobylitsa}}, \bibinfo {author} {\bibfnamefont {M.~V.}\ \bibnamefont {Polyakov}}, \bibinfo {author} {\bibfnamefont {I.}~\bibnamefont {Bornig}}, \bibinfo {author} {\bibfnamefont {K.}~\bibnamefont {Goeke}},\ and\ \bibinfo {author} {\bibfnamefont {C.}~\bibnamefont {Weiss}},\ }\href {https://doi.org/10.1103/PhysRevD.57.4325} {\bibfield  {journal} {\bibinfo  {journal} {Phys. Rev. D}\ }\textbf {\bibinfo {volume} {57}},\ \bibinfo {pages} {4325} (\bibinfo {year} {1998})},\ \Eprint {https://arxiv.org/abs/hep-ph/9710270} {arXiv:hep-ph/9710270} \BibitemShut {NoStop}%
\bibitem [{\citenamefont {Penttinen}\ \emph {et~al.}(2000)\citenamefont {Penttinen}, \citenamefont {Polyakov},\ and\ \citenamefont {Goeke}}]{Penttinen:1999th}%
  \BibitemOpen
  \bibfield  {author} {\bibinfo {author} {\bibfnamefont {M.}~\bibnamefont {Penttinen}}, \bibinfo {author} {\bibfnamefont {M.~V.}\ \bibnamefont {Polyakov}},\ and\ \bibinfo {author} {\bibfnamefont {K.}~\bibnamefont {Goeke}},\ }\href {https://doi.org/10.1103/PhysRevD.62.014024} {\bibfield  {journal} {\bibinfo  {journal} {Phys. Rev. D}\ }\textbf {\bibinfo {volume} {62}},\ \bibinfo {pages} {014024} (\bibinfo {year} {2000})},\ \Eprint {https://arxiv.org/abs/hep-ph/9909489} {arXiv:hep-ph/9909489} \BibitemShut {NoStop}%
\bibitem [{\citenamefont {Ossmann}\ \emph {et~al.}(2005)\citenamefont {Ossmann}, \citenamefont {Polyakov}, \citenamefont {Schweitzer}, \citenamefont {Urbano},\ and\ \citenamefont {Goeke}}]{Ossmann:2004bp}%
  \BibitemOpen
  \bibfield  {author} {\bibinfo {author} {\bibfnamefont {J.}~\bibnamefont {Ossmann}}, \bibinfo {author} {\bibfnamefont {M.~V.}\ \bibnamefont {Polyakov}}, \bibinfo {author} {\bibfnamefont {P.}~\bibnamefont {Schweitzer}}, \bibinfo {author} {\bibfnamefont {D.}~\bibnamefont {Urbano}},\ and\ \bibinfo {author} {\bibfnamefont {K.}~\bibnamefont {Goeke}},\ }\href {https://doi.org/10.1103/PhysRevD.71.034011} {\bibfield  {journal} {\bibinfo  {journal} {Phys. Rev. D}\ }\textbf {\bibinfo {volume} {71}},\ \bibinfo {pages} {034011} (\bibinfo {year} {2005})},\ \Eprint {https://arxiv.org/abs/hep-ph/0411172} {arXiv:hep-ph/0411172} \BibitemShut {NoStop}%
\bibitem [{\citenamefont {Wakamatsu}\ and\ \citenamefont {Tsujimoto}(2005)}]{Wakamatsu:2005vk}%
  \BibitemOpen
  \bibfield  {author} {\bibinfo {author} {\bibfnamefont {M.}~\bibnamefont {Wakamatsu}}\ and\ \bibinfo {author} {\bibfnamefont {H.}~\bibnamefont {Tsujimoto}},\ }\href {https://doi.org/10.1103/PhysRevD.71.074001} {\bibfield  {journal} {\bibinfo  {journal} {Phys. Rev. D}\ }\textbf {\bibinfo {volume} {71}},\ \bibinfo {pages} {074001} (\bibinfo {year} {2005})},\ \Eprint {https://arxiv.org/abs/hep-ph/0502030} {arXiv:hep-ph/0502030} \BibitemShut {NoStop}%
\bibitem [{\citenamefont {Kim}\ and\ \citenamefont {Weiss}(2024)}]{Kim:2023pll}%
  \BibitemOpen
  \bibfield  {author} {\bibinfo {author} {\bibfnamefont {J.-Y.}\ \bibnamefont {Kim}}\ and\ \bibinfo {author} {\bibfnamefont {C.}~\bibnamefont {Weiss}},\ }\href {https://doi.org/10.1016/j.physletb.2023.138387} {\bibfield  {journal} {\bibinfo  {journal} {Phys. Lett. B}\ }\textbf {\bibinfo {volume} {848}},\ \bibinfo {pages} {138387} (\bibinfo {year} {2024})},\ \Eprint {https://arxiv.org/abs/2310.16890} {arXiv:2310.16890 [hep-ph]} \BibitemShut {NoStop}%
\bibitem [{\citenamefont {Kim}\ and\ \citenamefont {Weiss}(2025)}]{Kim:2024ibz}%
  \BibitemOpen
  \bibfield  {author} {\bibinfo {author} {\bibfnamefont {J.-Y.}\ \bibnamefont {Kim}}\ and\ \bibinfo {author} {\bibfnamefont {C.}~\bibnamefont {Weiss}},\ }\href {https://doi.org/10.1103/PhysRevD.111.074007} {\bibfield  {journal} {\bibinfo  {journal} {Phys. Rev. D}\ }\textbf {\bibinfo {volume} {111}},\ \bibinfo {pages} {074007} (\bibinfo {year} {2025})},\ \Eprint {https://arxiv.org/abs/2411.17634} {arXiv:2411.17634 [hep-ph]} \BibitemShut {NoStop}%
\bibitem [{\citenamefont {Kim}(2025)}]{Kim:2025mol}%
  \BibitemOpen
  \bibfield  {author} {\bibinfo {author} {\bibfnamefont {J.-Y.}\ \bibnamefont {Kim}},\ }\href {https://doi.org/10.1103/qpzb-1skm} {\bibfield  {journal} {\bibinfo  {journal} {Phys. Rev. D}\ }\textbf {\bibinfo {volume} {112}},\ \bibinfo {pages} {074032} (\bibinfo {year} {2025})},\ \Eprint {https://arxiv.org/abs/2506.21013} {arXiv:2506.21013 [hep-ph]} \BibitemShut {NoStop}%
\bibitem [{\citenamefont {Minamiguchi}(2026)}]{Minamiguchi:2026qcb}%
  \BibitemOpen
  \bibfield  {author} {\bibinfo {author} {\bibfnamefont {J.}~\bibnamefont {Minamiguchi}}\ }\href {https://doi.org/10.1103/8wrp-hkth} {10.1103/8wrp-hkth} (\bibinfo {year} {2026}),\ \Eprint {https://arxiv.org/abs/2609.37354} {arXiv:2609.37354 [hep-ph]} \BibitemShut {NoStop}%
\bibitem [{\citenamefont {Pobylitsa}(2003)}]{Pobylitsa:2003ty}%
  \BibitemOpen
  \bibfield  {author} {\bibinfo {author} {\bibfnamefont {P.~V.}\ \bibnamefont {Pobylitsa}},\ }\href@noop {} {\  (\bibinfo {year} {2003})},\ \Eprint {https://arxiv.org/abs/hep-ph/0301236} {arXiv:hep-ph/0301236} \BibitemShut {NoStop}%
\bibitem [{\citenamefont {Wakamatsu}(2009)}]{Wakamatsu:2009fn}%
  \BibitemOpen
  \bibfield  {author} {\bibinfo {author} {\bibfnamefont {M.}~\bibnamefont {Wakamatsu}},\ }\href {https://doi.org/10.1103/PhysRevD.79.094028} {\bibfield  {journal} {\bibinfo  {journal} {Phys. Rev. D}\ }\textbf {\bibinfo {volume} {79}},\ \bibinfo {pages} {094028} (\bibinfo {year} {2009})},\ \Eprint {https://arxiv.org/abs/0903.1886} {arXiv:0903.1886 [hep-ph]} \BibitemShut {NoStop}%
\bibitem [{\citenamefont {Schweitzer}\ \emph {et~al.}(2013)\citenamefont {Schweitzer}, \citenamefont {Strikman},\ and\ \citenamefont {Weiss}}]{Schweitzer:2012hh}%
  \BibitemOpen
  \bibfield  {author} {\bibinfo {author} {\bibfnamefont {P.}~\bibnamefont {Schweitzer}}, \bibinfo {author} {\bibfnamefont {M.}~\bibnamefont {Strikman}},\ and\ \bibinfo {author} {\bibfnamefont {C.}~\bibnamefont {Weiss}},\ }\href {https://doi.org/10.1007/JHEP01(2013)163} {\bibfield  {journal} {\bibinfo  {journal} {JHEP}\ }\textbf {\bibinfo {volume} {01}},\ \bibinfo {pages} {163}},\ \Eprint {https://arxiv.org/abs/1210.1267} {arXiv:1210.1267 [hep-ph]} \BibitemShut {NoStop}%
\bibitem [{\citenamefont {Minamiguchi}\ and\ \citenamefont {Uji}(2026{\natexlab{b}})}]{Minamiguchi:2026usn}%
  \BibitemOpen
  \bibfield  {author} {\bibinfo {author} {\bibfnamefont {J.}~\bibnamefont {Minamiguchi}}\ and\ \bibinfo {author} {\bibfnamefont {T.}~\bibnamefont {Uji}},\ }\href@noop {} {\  (\bibinfo {year} {2026}{\natexlab{b}})},\ \Eprint {https://arxiv.org/abs/2610.39128} {arXiv:2610.39128 [hep-ph]} \BibitemShut {NoStop}%
\bibitem [{\citenamefont {Suh}\ \emph {et~al.}(2022)\citenamefont {Suh}, \citenamefont {Kim}, \citenamefont {Yang},\ and\ \citenamefont {Kim}}]{Suh:2022atr}%
  \BibitemOpen
  \bibfield  {author} {\bibinfo {author} {\bibfnamefont {J.-M.}\ \bibnamefont {Suh}}, \bibinfo {author} {\bibfnamefont {J.-Y.}\ \bibnamefont {Kim}}, \bibinfo {author} {\bibfnamefont {G.-S.}\ \bibnamefont {Yang}},\ and\ \bibinfo {author} {\bibfnamefont {H.-C.}\ \bibnamefont {Kim}},\ }\href {https://doi.org/10.1103/PhysRevD.106.054032} {\bibfield  {journal} {\bibinfo  {journal} {Phys. Rev. D}\ }\textbf {\bibinfo {volume} {106}},\ \bibinfo {pages} {054032} (\bibinfo {year} {2022})},\ \Eprint {https://arxiv.org/abs/2208.04447} {arXiv:2208.04447 [hep-ph]} \BibitemShut {NoStop}%
\bibitem [{\citenamefont {Silva}\ \emph {et~al.}(2005)\citenamefont {Silva}, \citenamefont {Kim}, \citenamefont {Urbano},\ and\ \citenamefont {Goeke}}]{Silva:2005fa}%
  \BibitemOpen
  \bibfield  {author} {\bibinfo {author} {\bibfnamefont {A.}~\bibnamefont {Silva}}, \bibinfo {author} {\bibfnamefont {H.-C.}\ \bibnamefont {Kim}}, \bibinfo {author} {\bibfnamefont {D.}~\bibnamefont {Urbano}},\ and\ \bibinfo {author} {\bibfnamefont {K.}~\bibnamefont {Goeke}},\ }\href {https://doi.org/10.1103/PhysRevD.72.094011} {\bibfield  {journal} {\bibinfo  {journal} {Phys. Rev. D}\ }\textbf {\bibinfo {volume} {72}},\ \bibinfo {pages} {094011} (\bibinfo {year} {2005})},\ \Eprint {https://arxiv.org/abs/hep-ph/0509281} {arXiv:hep-ph/0509281} \BibitemShut {NoStop}%
\bibitem [{\citenamefont {Lorce}\ \emph {et~al.}(2011)\citenamefont {Lorce}, \citenamefont {Pasquini},\ and\ \citenamefont {Vanderhaeghen}}]{Lorce:2011dv}%
  \BibitemOpen
  \bibfield  {author} {\bibinfo {author} {\bibfnamefont {C.}~\bibnamefont {Lorce}}, \bibinfo {author} {\bibfnamefont {B.}~\bibnamefont {Pasquini}},\ and\ \bibinfo {author} {\bibfnamefont {M.}~\bibnamefont {Vanderhaeghen}},\ }\href {https://doi.org/10.1007/JHEP05(2011)041} {\bibfield  {journal} {\bibinfo  {journal} {JHEP}\ }\textbf {\bibinfo {volume} {05}},\ \bibinfo {pages} {041}},\ \Eprint {https://arxiv.org/abs/1102.4704} {arXiv:1102.4704 [hep-ph]} \BibitemShut {NoStop}%
\bibitem [{\citenamefont {Di~Vecchia}\ and\ \citenamefont {Veneziano}(1980)}]{DiVecchia:1980yfw}%
  \BibitemOpen
  \bibfield  {author} {\bibinfo {author} {\bibfnamefont {P.}~\bibnamefont {Di~Vecchia}}\ and\ \bibinfo {author} {\bibfnamefont {G.}~\bibnamefont {Veneziano}},\ }\href {https://doi.org/10.1016/0550-3213(80)90370-3} {\bibfield  {journal} {\bibinfo  {journal} {Nucl. Phys. B}\ }\textbf {\bibinfo {volume} {171}},\ \bibinfo {pages} {253} (\bibinfo {year} {1980})}\BibitemShut {NoStop}%
\bibitem [{\citenamefont {Witten}(1980)}]{Witten:1980sp}%
  \BibitemOpen
  \bibfield  {author} {\bibinfo {author} {\bibfnamefont {E.}~\bibnamefont {Witten}},\ }\href {https://doi.org/10.1016/0003-4916(80)90325-5} {\bibfield  {journal} {\bibinfo  {journal} {Annals Phys.}\ }\textbf {\bibinfo {volume} {128}},\ \bibinfo {pages} {363} (\bibinfo {year} {1980})}\BibitemShut {NoStop}%
\bibitem [{\citenamefont {Wakamatsu}\ and\ \citenamefont {Yoshiki}(1991)}]{Wakamatsu:1990ud}%
  \BibitemOpen
  \bibfield  {author} {\bibinfo {author} {\bibfnamefont {M.}~\bibnamefont {Wakamatsu}}\ and\ \bibinfo {author} {\bibfnamefont {H.}~\bibnamefont {Yoshiki}},\ }\href {https://doi.org/10.1016/0375-9474(91)90263-6} {\bibfield  {journal} {\bibinfo  {journal} {Nucl. Phys. A}\ }\textbf {\bibinfo {volume} {524}},\ \bibinfo {pages} {561} (\bibinfo {year} {1991})}\BibitemShut {NoStop}%
\bibitem [{\citenamefont {Jain}\ \emph {et~al.}(1988)\citenamefont {Jain}, \citenamefont {Johnson}, \citenamefont {Meissner}, \citenamefont {Park},\ and\ \citenamefont {Schechter}}]{Jain:1987sz}%
  \BibitemOpen
  \bibfield  {author} {\bibinfo {author} {\bibfnamefont {P.}~\bibnamefont {Jain}}, \bibinfo {author} {\bibfnamefont {R.}~\bibnamefont {Johnson}}, \bibinfo {author} {\bibfnamefont {U.~G.}\ \bibnamefont {Meissner}}, \bibinfo {author} {\bibfnamefont {N.~W.}\ \bibnamefont {Park}},\ and\ \bibinfo {author} {\bibfnamefont {J.}~\bibnamefont {Schechter}},\ }\href {https://doi.org/10.1103/PhysRevD.37.3252} {\bibfield  {journal} {\bibinfo  {journal} {Phys. Rev. D}\ }\textbf {\bibinfo {volume} {37}},\ \bibinfo {pages} {3252} (\bibinfo {year} {1988})}\BibitemShut {NoStop}%
\bibitem [{\citenamefont {Meissner}\ \emph {et~al.}(1989)\citenamefont {Meissner}, \citenamefont {Kaiser}, \citenamefont {Weigel},\ and\ \citenamefont {Schechter}}]{Meissner:1988iv}%
  \BibitemOpen
  \bibfield  {author} {\bibinfo {author} {\bibfnamefont {U.~G.}\ \bibnamefont {Meissner}}, \bibinfo {author} {\bibfnamefont {N.}~\bibnamefont {Kaiser}}, \bibinfo {author} {\bibfnamefont {H.}~\bibnamefont {Weigel}},\ and\ \bibinfo {author} {\bibfnamefont {J.}~\bibnamefont {Schechter}},\ }\href {https://doi.org/10.1103/PhysRevD.39.1956} {\bibfield  {journal} {\bibinfo  {journal} {Phys. Rev. D}\ }\textbf {\bibinfo {volume} {39}},\ \bibinfo {pages} {1956} (\bibinfo {year} {1989})},\ \bibinfo {note} {[Erratum: Phys.Rev.D 40, 262 (1989)]}\BibitemShut {NoStop}%
\bibitem [{\citenamefont {Schechter}\ and\ \citenamefont {Weigel}(1999)}]{Schechter:1999hg}%
  \BibitemOpen
  \bibfield  {author} {\bibinfo {author} {\bibfnamefont {J.}~\bibnamefont {Schechter}}\ and\ \bibinfo {author} {\bibfnamefont {H.}~\bibnamefont {Weigel}},\ }\href@noop {} {\ ,\ \bibinfo {pages} {337} (\bibinfo {year} {1999})},\ \Eprint {https://arxiv.org/abs/hep-ph/9907554} {arXiv:hep-ph/9907554} \BibitemShut {NoStop}%
\bibitem [{\citenamefont {Diehl}(2003)}]{Diehl:2003ny}%
  \BibitemOpen
  \bibfield  {author} {\bibinfo {author} {\bibfnamefont {M.}~\bibnamefont {Diehl}},\ }\href {https://doi.org/10.1016/j.physrep.2003.08.002} {\bibfield  {journal} {\bibinfo  {journal} {Phys. Rept.}\ }\textbf {\bibinfo {volume} {388}},\ \bibinfo {pages} {41} (\bibinfo {year} {2003})},\ \Eprint {https://arxiv.org/abs/hep-ph/0307382} {arXiv:hep-ph/0307382} \BibitemShut {NoStop}%
\bibitem [{\citenamefont {Kahana}\ and\ \citenamefont {Ripka}(1984)}]{Kahana:1984be}%
  \BibitemOpen
  \bibfield  {author} {\bibinfo {author} {\bibfnamefont {S.}~\bibnamefont {Kahana}}\ and\ \bibinfo {author} {\bibfnamefont {G.}~\bibnamefont {Ripka}},\ }\href {https://doi.org/10.1016/0375-9474(84)90692-4} {\bibfield  {journal} {\bibinfo  {journal} {Nucl. Phys. A}\ }\textbf {\bibinfo {volume} {429}},\ \bibinfo {pages} {462} (\bibinfo {year} {1984})}\BibitemShut {NoStop}%
\bibitem [{\citenamefont {Kubota}\ \emph {et~al.}(1999)\citenamefont {Kubota}, \citenamefont {Wakamatsu},\ and\ \citenamefont {Watabe}}]{Kubota:1999hx}%
  \BibitemOpen
  \bibfield  {author} {\bibinfo {author} {\bibfnamefont {T.}~\bibnamefont {Kubota}}, \bibinfo {author} {\bibfnamefont {M.}~\bibnamefont {Wakamatsu}},\ and\ \bibinfo {author} {\bibfnamefont {T.}~\bibnamefont {Watabe}},\ }\href {https://doi.org/10.1103/PhysRevD.60.014016} {\bibfield  {journal} {\bibinfo  {journal} {Phys. Rev. D}\ }\textbf {\bibinfo {volume} {60}},\ \bibinfo {pages} {014016} (\bibinfo {year} {1999})},\ \Eprint {https://arxiv.org/abs/hep-ph/9902329} {arXiv:hep-ph/9902329} \BibitemShut {NoStop}%
\bibitem [{\citenamefont {Weigel}\ and\ \citenamefont {Takyi}(2021)}]{Weigel:2021pbr}%
  \BibitemOpen
  \bibfield  {author} {\bibinfo {author} {\bibfnamefont {H.}~\bibnamefont {Weigel}}\ and\ \bibinfo {author} {\bibfnamefont {I.}~\bibnamefont {Takyi}},\ }\href {https://doi.org/10.3390/sym13010108} {\bibfield  {journal} {\bibinfo  {journal} {Symmetry}\ }\textbf {\bibinfo {volume} {13}},\ \bibinfo {pages} {108} (\bibinfo {year} {2021})},\ \Eprint {https://arxiv.org/abs/2101.02114} {arXiv:2101.02114 [hep-ph]} \BibitemShut {NoStop}%
\end{thebibliography}%

\end{document}